\documentstyle[equations]{mn}

\newlength{\colwidthf}
\newlength{\hcolwidthf}
\newlength{\colwidth}
\newlength{\hcolwidth}
\input psfig
\newif\ifpsfiles\psfilestrue
\def\getfig#1#2{\ifpsfiles\psfig{figure=#1,width=\hsize}\else\vskip#2\fi}

\font\ninerm=cmr9    \font\sixrm=cmr5
  
\font\nineit=cmti9  
\font\ninesl=cmsl9
\font\ninei=cmmi9    \font\sixi=cmmi5
\skewchar\ninei='177
\font\ninesy=cmsy9  \font\sixsy=cmsy5
\skewchar\ninesy='60
\font\ninebf=cmbx9  \font\sixbf=cmbx5
\font\nineex=cmex10 scaled 833

\font\ninett=cmtt9
\def\adjustlinespace{\baselineskip=\baselineskip}
\def\ninepoint{\textfont0=\ninerm \scriptfont0=\sixrm 
                \def\rm{\fam0\ninerm}\relax
                \textfont1=\ninei \scriptfont1=\sixi 
                \def\mit{\fam1}\def\oldstyle{\fam1\ninei}\relax
                \textfont2=\ninesy \scriptfont2=\sixsy 
                \def\cal{\fam2}\relax
                \textfont3=\nineex \scriptfont3=\nineex 
                \def\it{\fam\itfam\nineit}\relax
                \textfont\itfam=\nineit
                \def\sl{\fam\slfam\ninesl}\relax
                \textfont\slfam=\ninesl
                \def\bf{\fam\bffam\ninebf}\relax
                \textfont\bffam=\ninebf \scriptfont\bffam=\sixbf
                \def\tt{\fam\ttfam\ninett}\relax
                \textfont\ttfam=\ninett
                \setbox\strutbox=\hbox{\vrule
                     hnine7pt depth2pt width0pt}\baselineskip=9pt
                \adjustlinespace
                \rm}
\font\caps=cmcsc10			   
\def\etal{{\it et~al.\ }}
\def\aa #1 #2 {A\&A, #1, #2}
\def\aas #1 #2 {A\&AS, #1, #2}
\def\acm #1 #2 {ACM-Trans Math Software, #1, #2}
\def\ada #1 #2 {Ann Astrophys, #1, #2}
\def\agabstr #1 #2 {Astr Ges Abstr Ser, #1, #2}
\def\aj #1 #2 {AJ, #1, #2}
\def\anach #1 #2 {Astr Nachr, #1, #2}
\def\apj #1 #2 {ApJ, #1, #2}
\def\apjl #1 #2 {ApJL, #1, #2}
\def\apjs #1 #2 {ApJS, #1, #2}
\def\araa #1 #2 {ARAA, #1, #2}
\def\apss #1 #2 {ApSpaceS, #1, #2}
\def\celmech #1 #2 {Cel Mech, #1, #2}
\def\esom #1 #2 {ESO Messenger, #1, #2}
\def\fundcp #1 #2 {FunCosP, #1, #2}
\def\jcp #1 #2 {J Comp Phys, #1, #2}
\def\jfm #1 #2 {J Fluid Mech, #1, #2}
\def\jmp #1 #2 {J Math Phys, #1, #2}
\def\ma #1 #2 {Mitt Astr Ges, #1, #2}
\def\mn #1 #2 {MNRAS, #1, #2}
\def\nat #1 #2 {Nat, #1, #2}
\def\obs #1 #2 {Observatory, #1, #2}
\def\pasj #1 #2 {PASJ, #1, #2}
\def\pasp #1 #2 {PASP, #1, #2}
\def\phyr #1 #2 {PhysRep, #1, #2}
\def\physd #1 #2 {Physica D, #1, #2}
\def\rpp #1 #2 {RepProgPhys, #1, #2}
\def\ssr #1 #2 {Sp Sci Rev, #1, #2}
\def\iau127#1{in de Zeeuw P.T. ed, Structure and Dynamics of 
     Elliptical Galaxies, IAU Symp.~No.~127. Reidel, Dordrecht, p.~#1}

\def\inbook#1#2#3#4#5#6{in: #1%
\if#2-%
\else%
, #2%
\fi%
\if#3-%
\else%
, ed.\ #3%
\fi%
\if#5-%
 {\if#4-%
 \else,%
   (#4)%
 \fi}%
\else%
 {\if#4-%
, (#5)%
\else%
, (#5:#4)%
\fi}%
\fi%
\if#6-%
.%
\else%
, #6.%
\fi%
}

\def\spose#1{\hbox to 0pt{#1\hss}}
\def\lta{\mathrel{\spose{\lower 3pt\hbox{$\mathchar"218$}}
     \raise 2.0pt\hbox{$\mathchar"13C$}}}
\def\gta{\mathrel{\spose{\lower 3pt\hbox{$\mathchar"218$}}
     \raise 2.0pt\hbox{$\mathchar"13E$}}}

\def\=#1{\overline{#1}}

\def\d{{\rm d}}

\def\deg{^\circ}             
\def\MYM{{\mu\rm m}}

\def\kms{{\rm\,km\,s^{-1}}}

\def\pc{{\rm\,pc}}
\def\kpc{{\rm\,kpc}}

\def\msun{{\rm\,M_\odot}}

\def\gyr{{\rm\,Gyr}}
\def\MHZ{\,{\rm MHz}}

\def\rms{{\caps rms}}

\def\etal{et~al.\ }

\def\d{\rm d}

\def\KPC{\rm kpc}
\def\PC{\rm pc}
\def\KM{\rm km}
\def\SEC{\rm s}
\def\KMS{\KM \SEC^{-1}}
\def\KMSKPC{\KM \SEC^{-1} \KPC^{-1}}

\def\GYR{\rm Gyr}

\def\MSUN{M_\odot}

\def\msun{\MSUN}

\def\vek#1{{\bmath #1}}

\newcommand{\eq}[1]{
	\begin{equation}
	#1
	\end{equation}
}
\newcommand{\eqn}[2]{
	\begin{equation}
	#2
	\label{#1}
	\end{equation}
}

\def\spose#1{\hbox to 0pt{#1\hss}}
\def\lta{\mathrel{\spose{\lower 3pt\hbox{$\mathchar"218$}}
     \raise 2.0pt\hbox{$\mathchar"13C$}}}
\def\gta{\mathrel{\spose{\lower 3pt\hbox{$\mathchar"218$}}
     \raise 2.0pt\hbox{$\mathchar"13E$}}}

\def\=#1{\overline{#1}}

\def\twco{$^{12}$CO}
\def\thco{$^{13}$CO}

\def\lvplot{($l,v$) diagram}
\def\lvplots{\lvplot s}
\def\csnd{c_{\rm s}}
\def\PHIBAR{\varphi_{\rm bar}}
\def\phibar{\varphi_{\rm bar}}
\def\Phieff{\Phi_{\rm eff}}
\def\OmegaP{\Omega_{\rm P}}
\def\VSUN{V_0}

\def\MHZ{\,{\rm MHz}}

\def\sumlm{\sum_{l, m}}
\def\suml{\sum_l}

\def\sign{{\rm sign}}

\def\MYM{{\mu\rm m}}

\author[P.~Englmaier and O.E.~Gerhard]{Peter~Englmaier and Ortwin~Gerhard}

\title{Gas Dynamics and Large--Scale Morphology of the Milky Way Galaxy}

\begin{document}
\maketitle
\begin{abstract}
We present a new model for the gas dynamics in the galactic disk
inside the Sun's orbit. Quasi--equilibrium flow solutions are
determined in the gravitational potential of the deprojected COBE NIR
bar and disk, complemented by a central cusp
and, in some models, an outer halo. These models generically lead to
four--armed spiral structure between corotation of the bar and the
solar circle; their large--scale morphology is not sensitive to the
precise value of the bar's pattern speed, to the orientation of the
bar with respect to the observer, and to whether or not the spiral
arms carry mass.

Our best model provides a coherent interpretation of many observed gas
dynamical features. Its four-armed spiral structure outside corotation
reproduces quantitatively the directions to the five main spiral arm
tangents at $\vert l\vert \leq 60\deg$ observed in a variety of
tracers. The 3-kpc-arm is identified with one of the model arms
emanating from the ends of the bar, extending into the corotation
region. The model features an inner gas disk with a cusped orbit
shock transition to an $x_2$ orbit disk of radius $R\sim 150 \pc$.

The bar's corotation radius is fairly well--constrained at $R_c\simeq
3.5\pm 0.5 \kpc$. The best value for the orientation angle of the bar
is probably $20-25\deg$, but the uncertainty is large since no
detailed quantitative fit to all features in the observed \lvplot s is
yet possible.  The Galactic terminal velocity curve from HI and CO
observations out to $l\simeq \pm45\deg$ ($\sim 5\kpc$) is
approximately described by a maximal disk model with constant
mass--to--light ratio for the NIR bulge and disk.

\end{abstract}
%
%
\section{Introduction}

Although the Milky Way is in many ways the best studied example of a
disk galaxy, it has proven exceedingly difficult to reliably determine
its large--scale properties, such as the overall morphology, the
structural parameters of the main components, the spiral arm pattern,
and the shape of the Galactic rotation curve. A large part of this
difficulty is due to distance ambiguities and to the unfortunate
location of the solar system within the Galactic dust layer, which
obscures the stellar components of the Galaxy in the optical
wavebands. With the advent of comprehensive near--infrared
observations by the COBE/DIRBE satellite and other ground- and
space--based experiments, this situation has improved dramatically.
These data offer a new route to mapping out the Galaxy's stellar
components, to connecting their gravitational potential with the
available gas and stellar kinematic observations, and thereby to
understanding the large--scale structure and dynamics of the Milky Way
Galaxy.

From radio and mm--observations it has long been known that the
atomic and molecular gas in the inner Galaxy does not move quietly on
circular orbits: ``forbidden'' and non--circular motions in excess of
$100\kms$ are seen in longitude--velocity ($l,v$)--diagrams (e.g.,
Burton \& Liszt 1978, Dame \etal 1987, Bally \etal 1987).  Some of the
more prominent features indicating non--circular motions are the
$3\kpc$--arm, the molecular parallelogram (``expanding molecular
ring''), and the unusually high central peak in the terminal velocity
curve at $l\simeq \pm 2\deg$.  Many papers in the past have suggested
that these forbidden velocities are best explained if one assumes that
the gas moves on elliptical orbits in a barred gravitational potential
(Peters 1975, Cohen \& Few 1976, Liszt \& Burton 1980, Gerhard \&
Vietri 1986, Mulder \& Liem 1986, Binney \etal 1991, Wada \etal 1994).

In the past few years, independent evidence for a bar in the inner
Galaxy has been mounting from NIR photometry (Blitz \& Spergel 1991,
Weiland \etal 1994, Dwek \etal 1995), from IRAS and clump giant source
counts (Nakada \etal 1991, Whitelock \& Catchpole 1992, Nikolaev \&
Weinberg 1997, Stanek \etal 1997), from the measured large
microlensing optical depth towards the bulge (Paczynski \etal 1994,
Zhao, Rich \& Spergel 1996) and possibly also from stellar kinematics
(Zhao, Spergel \& Rich 1994).  See Gerhard (1996) and Kuijken (1996)
for recent reviews.

The currently best models for the distribution of old stars in the
inner Galaxy are based on the NIR data from the DIRBE experiment on
COBE. Because extinction is important towards the Galactic nuclear
bulge even at $2\,\mu$m, the DIRBE data must first be corrected (or
`cleaned') for the effects of extinction. This has been done by
Spergel, Malhotra \& Blitz (1996), using a fully three-dimensional
model of the dust distribution (see also Freudenreich 1998).  Binney,
Gerhard \& Spergel (1997, hereafter BGS) used a Richardson--Lucy
algorithm to fit a non-parametric model of $j({\bf r})$ to the cleaned
data of Spergel \etal under the assumption of eight-fold (triaxial)
symmetry with respect to three mutually orthogonal planes.  When the
orientation of the symmetry planes is fixed, the recovered emissivity
$j({\bf r})$ appears to be essentialy unique (see also Binney \&
Gerhard 1996, Bissantz \etal 1997), but physical models matching the
DIRBE data can be found for a range of bar orientation angles, $15\deg
\lta \PHIBAR \lta 35\deg$ (BGS). $\PHIBAR$ measures the angle in the
Galactic plane between the bar's major axis at $l>0$ and the
Sun--centre line. For the favoured $\PHIBAR=20\deg$, the deprojected
luminosity distribution shows an elongated bulge with axis ratios
10:6:4 and semi--major axis $\sim 2\kpc$, surrounded by an elliptical
disk that extends to $\sim 3.5\kpc$ on the major axis and $\sim 2\kpc$
on the minor axis.

Outside the bar, the NIR luminosity distribution shows a maximum in
the emissivity $\sim 3\kpc$ down the minor axis, which corresponds to
the ring--like structure discussed by Kent, Dame \& Fazio (1991), and
which might well be due to incorrectly deprojected strong spiral arms.
From the study of HII-Regions, molecular clouds and the Galactic
magnetic field it appears that the Milky Way may have four main spiral
arms (Georgelin \& Georgelin 1976, Caswell \& Heynes 1987; Sanders,
Scoville \& Solomon 1985, Grabelsky \etal 1988; Vall\'ee 1995).  These
are located outside the $3\kpc$-arm in the \lvplot\ and are probably
related to the so--called molecular ring (Dame 1993), although it is
unclear precisely how.  One problem with this is that the
distances to the tracers used to map out the spiral arms are usually
computed on the basis of a circular gas flow model.  The errors due to
this assumption are not likely to be large, but cannot be reliably
assessed until more realistic gas flow models including non--circular
motions are available.

The goal of this paper is to construct gas-dynamical models for the
inner Milky Way that connect the Galactic bar/bulge and disk as
observed in the COBE NIR luminosity distribution with the kinematic
observations of HI and molecular gas in the \lvplot. In this way we
hope to constrain parameters like the orientation, mass, and pattern
speed of the bar, and to reach a qualitative understanding of the
Galactic spiral arms and other main features in the observed \lvplot s.

In barred potentials, gas far from resonances settles on periodic
orbits such as those of the $x_1$- and $x_2$-orbit families, and some
important aspects of the gas flow can be understood by considering the
closed periodic orbits (e.g., Binney \etal 1991). However, near
transitions of the gas between orbit families, along the leading edges
of the bar, and in spiral arms, shocks form in the gas flow which can
only be studied by gas dynamical simulations (e.g., Roberts, van
Albada \& Huntley 1979, Athanassoula 1992b, Englmaier \& Gerhard
1997).

In this paper, we use the Smoothed Particle Hydrodynamics (SPH) method
to study the gas flow in the gravitational potential of the
non-parametrically deprojected COBE/DIRBE light distribution of
Binney, Gerhard \& Spergel (1997), assuming a constant mass to NIR
luminosity ratio. The gas settles to an approximately
quasi--stationary flow, and the resulting model \lvplot s enable us to
understand many aspects of the observations of HI and molecular gas.

The paper is organized as follows. In Section 2 we give a brief review
of the main observational constraints, followed in Section 3 by a
description of our models for the mass distribution and gravitational
potential, and for the treatment of the gas. Section 4 describes the
results from a sequence of gas dynamical models designed to constrain
the most important parameters, and compares these models with
observations. Our results and conclusions are summarized in Section 5.


\section{Summary of observational constraints}

\subsection{Surveys and \lvplot s}

The kinematics and distribution of gas in the Milky Way disk have been
mapped in 21-cm neutral hydrogen emission and in mm-line emission from
molecular species, most notably \twco.  HI surveys include Burton
(1985), Kerr \etal (1986), Stark \etal (1992) and Hartmann \& Burton
(1997), surveys in \twco\ include Sanders \etal (1986), Dame \etal
(1987) and Stark \etal (1988), and the most comprehensive survey in
\thco\ is that of Bally \etal (1987). The large scale morphology of
the gas based on these data is discussed in the review by Burton
(1992).

\begin{figure*}
\ifpsfiles\psfig{figure=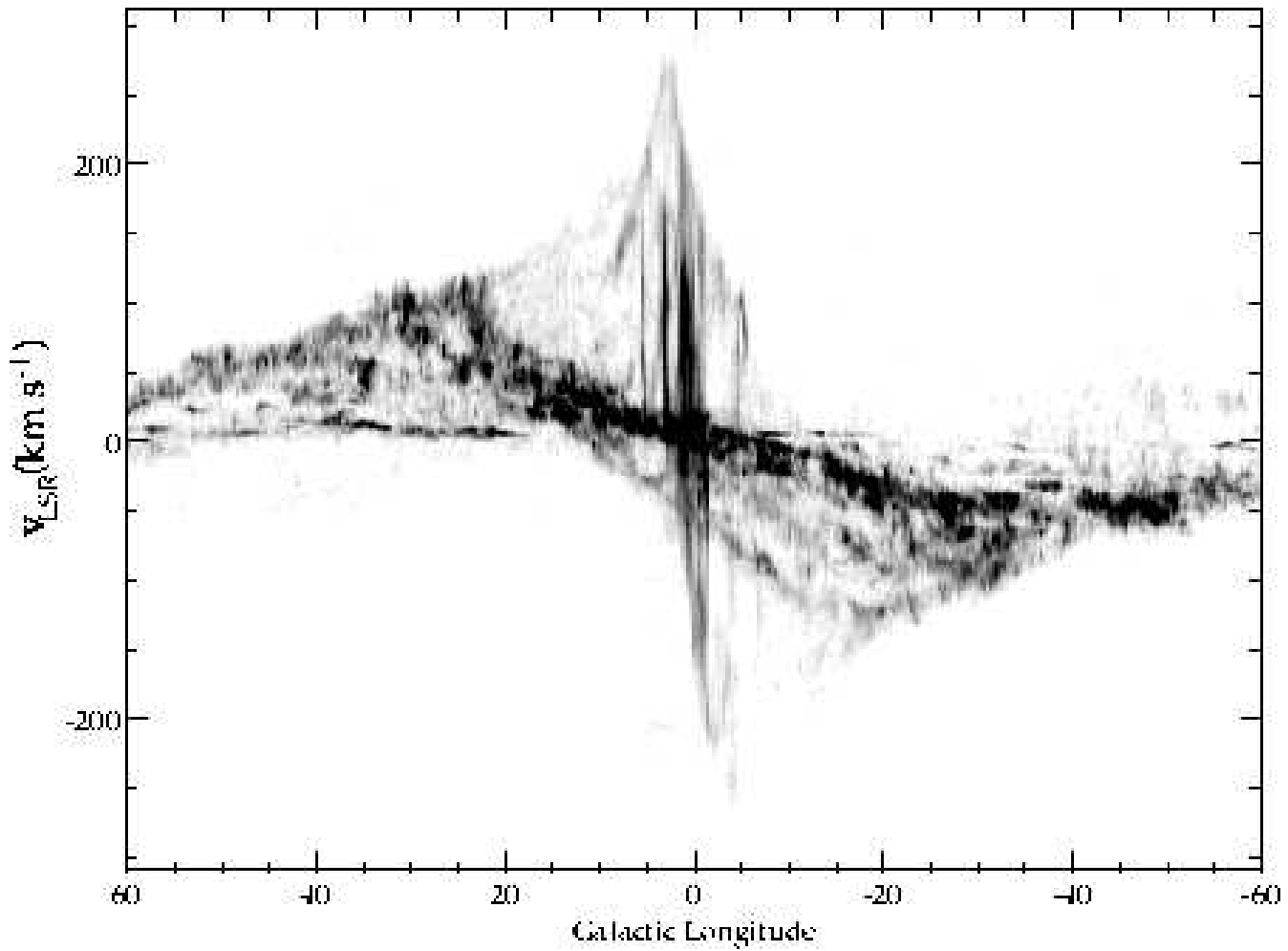,width=\hsize}
\else\vskip6cm\fi
\caption[]{%
\lvplot\ for \twco\ from unpublished data by Dame \etal (1998). This
figure contains all emission integrated over latitudes between
$b=-2\deg$ and $b=2\deg$.  The grey scale is adjusted such as to
emphasize spiral arm structures.
}
\label{obslvplot}
\end{figure*}

These surveys show a complicated distribution of gas in
longitude--latitude--radial velocity $(l,b,v_r)$ space or, integrating
over some range in latitude, in the so--called \lvplot.
Fig.~\ref{obslvplot} shows an \lvplot\ observed in \twco\ by Dame
\etal (1998, see also Dame \etal 1987) and transformed by them to the
LSR frame by subtracting $\vert\bmath{v_\odot}\vert=20\kms$ towards
$(l,b)=(56.2^\circ,22.8^\circ)$ for the motion of the Sun with respect
to the LSR (This corresponds to inward radial and forward components
of the solar motion of $u_\odot=-10.3 \kms$ and $v_\odot=15.3 \kms$.)
The general morphology of the \lvplot\ in \twco\ is broadly similar to
that obtained from HI 21-cm emission (see, e.g., Burton 1992).

In general, no additional distance information for the gas is
available; thus -- unlike in the case of external galaxies -- the
spatial distribution of galactic gas cannot be directly
inferred. Converting line--of--sight velocities to distances, on the
other hand, requires a model of the gas flow.

\subsection{Interpretation of \lvplot s}

For the comparison of observed and model \lvplot s, it is useful to
first consider an axisymmetric disk with gas in circular rotation
(e.g., Mihalas \& Binney 1981). In this model, an observer on a
circular orbit will find the following results:

  (i) Gas on the same circular orbit as the observer will have zero
relative radial velocity.

  (ii) For clouds on a different circular
orbit with velocity $v(R)$, the measured radial velocity is
\begin{equation}
v_r=(\omega-\omega_0) R_0 \sin l,
\label{circvterm}
\end{equation}
 where $\omega(R)\equiv v(R)/R$ is the angular rotation rate, $R_0$ is
the galactocentric radius of the observer, and the index $0$ to a
function or variable denotes its value at $R_0$.  From this equation
we see that, as long as $\omega(R)$ decreases outwards, the radial
velocities have the same sign as $\sin l$ for gas inside the observer,
and the opposite sign for gas on circular orbits outside the observer.

(iii) For clouds inside the observers orbit ($-90<l<90\deg$), the
maximal radial velocity along a given line--of--sight $l$ is $v_r =
\sign(l)v(R)-\VSUN \sin l$. This so--called terminal velocity is
reached at the tangent point to the circular orbit with $R=R_0 \sin
l$. For $l>0$ ($l<0$) $v_r$ increases (decreases) from zero for clouds
near the observer to the terminal velocity at a distance corresponding
to the circular orbit with $R=R_0 \sin l$; it then decreases
(increases) again with distance from the observer and changes sign
when crossing the observer's radius at the far side of the Galaxy. The
terminal velocities define the upper (lower) envelope in the \lvplot\
for $0<l<90\deg$ ($-90\deg<l<0$).

(iv) A circular orbit of given radius follows a sinusoidal path in the
\lvplot [cf.\ equation~(1)], within a longitude range bounded
by $l=\pm \arcsin(R/R_0)$. In the inner Galaxy ($-45\deg\lta l\lta
45\deg$) circular orbits thus approximately trace out straight lines
through the orgin in the \lvplot.

(v) The edge of the Galaxy (the outermost circular orbit) results in a
sine shaped envelope in the \lvplot\ with negative radial velocities
for positive longitudes ($0\deg<l<180^\circ$) and vice versa. The
sign of $v_r$ on this envelope is different from that on the terminal
velocity envelope.

(vi) For a circular orbit model, one can derive the rotation curve of
the inner Galaxy from the observed terminal velocities and
eq.~(\ref{circvterm}).  This requires knowledge of the galactocentric
radius $R_0$ and rotation velocity $\VSUN$ of the local standard of
rest (LSR) as well as the motion of the Sun with respect to the LSR.

Fig.~\ref{obslvplot} and similar HI 21 cm \lvplot s (see, e.g., Burton
1992) show gas between $90\deg$ and $-90\deg$ longitude whose radial
velocities have the wrong sign for it to be gas on circular orbits
inside the solar radius. Yet this gas is evidently connected to other
gas in the inner Galaxy and is not associated with gas from outside
the solar radius. These so--called forbidden velocities, up to
$v_r\simeq 100\kms$, are a direct signature of non--circular orbits in
the inner Galaxy, and they have been the basis of previous
interpretations of the inner Galaxy gas flow in terms of a rotating
bar potential (e.g., Peters 1975, Mulder \& Liem 1986, Binney \etal
1991).

One of the most prominent such features is the so--called $3
\kpc$--arm, visible in Fig.~\ref{obslvplot} as the dense ridge of
emission extending from $(l\simeq10\deg,v=0)$ through $(l=0,v\simeq
-50\kms)$ to $(l\simeq -22\deg,v\simeq-120\kms)$.

At longitudes
$|l|\gta 25\deg$, the circular orbit model {\sl is} a reasonable
description of the observed gas kinematics.  Most of the emission in
\twco\ in fact comes from a gas annulus between about 4 and $7\,\KPC$
galactocentric radius, the so--called molecular ring (e.g., Dame
1993), which probably consists of two pairs of tightly wound spiral
arms (see Section 4). A detailed interpretation of spiral arms in the
\lvplot\ requires a full gas dynamical model, as high intensities in
Fig.~\ref{obslvplot} can be due to either high intrinsic gas densities
or due to velocity crowding.

\subsection{Terminal velocities}


The top left and bottom right envelopes on the \lvplot\ in
Fig.~\ref{obslvplot} mark the terminal velocities. The terminal
velocity curves will be used below for comparing with different models
and for calibrating the models' mass--to--light ratios. As will be
seen in \S\ref{sec-termcurve} below, the terminal velocities in \twco\
and HI 21cm and between different surveys agree to a precision of
$\sim 10\kms$ in most places, but there are some regions with larger
discrepancies.

Of particular interest is the strong peak in the terminal velocity
curve with $v_t\simeq 260\kms$ at $l\simeq 2\deg$. Outwards from
there the drop in $v_t$ is very rapid; for a constant mass--to--infrared
luminosity ratio it would be nearly Keplerian and would be hard to
reproduce in an axisymmetric bulge model (Kent 1992). Instead, the
rapid drop is probably connected with a change of orbit shape over
this region (Gerhard \& Vietri 1986). In the model of Binney \etal (1991),
the peak is associated with the cusped orbit in the rotating barred
potential, and the subsequent drop of $v_t$ with $l$ reflects the
shapes of the adjacent $x_1$--orbits.

\subsection{Spiral arms}
\label{sec-obsarms}

From distant galaxies we know that spiral arms are traced by molecular
gas emission.  Indeed one can identify some of the dense emission
ridges in Fig.~\ref{obslvplot} with Galactic spiral arms; where these
meet the terminal velocity curve, they can be recognized as `bumps'
where $\partial v_t / \partial l \simeq 0$. In addition, spiral arms
are clearly visible in the distribution of various tracers, such as
HII regions.

\begin{figure}
\getfig{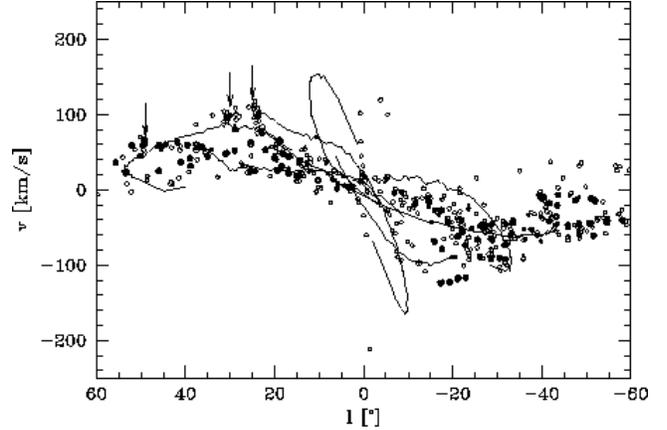}{8.2cm}
\caption[]{%
\lvplot\ for HII regions from Georgelin \& Georgelin (1976),
Downes \etal (1980) and Caswell \& Haynes (1987) (open
circles); and for massive molecular clouds from Dame \etal (1986) and
Bronfman (1989) (filled circles). The arrows point to the positions of
the densest clusters of warm CO clouds along the terminal curve, which
are presumably located in spiral arm shocks (Solomon, Sanders \&
Rivolo 1985). For the sake of clarity, clouds with less than
$10^{5.5}\,\MSUN$ have been omitted from Bronfman's data, as well as
clouds within the smallest brightness bin from the Georgelin \&
Georgelin sample. For illustration, the thin lines show the locations of
the gas spiral arms in our standard $\phibar=20\deg$ COBE bulge and disk
model without dark halo; cf.~Section 4.
}
\label{lvtracers}
\end{figure}

Fig.~\ref{lvtracers} shows an \lvplot\ of several classes of objects
which are useful as discrete tracers of dense gas in spiral arms. On
each side, we can identify two spiral arm tangents at around
$\pm\sim30$ and $\pm\sim50^\circ$. On the northern side, the
$\sim30^\circ$ component splits up into two components at $\sim30\deg$ and
$\sim25^\circ$ longitude, which is also evident from the Solomon \etal
(1985) data.

Table 1 lists a number of tracers that have been used to delineate
spiral arms. The inferred spiral arm tangents coincide with features
along the terminal curve in Fig.~\ref{obslvplot}; compare also
Fig.~\ref{obslvplot} and Fig.~\ref{lvtracers}. In the inner Galaxy
there are thus five main arm tangents, of which the Scutum tangent is
double in a number of tracers. The inner Scutum tangent at $l\simeq
25\deg$ is sometimes referred to as the northern 3--kpc arm.

While the main spiral arm tangents on both sides of the Galactic
center are thus fairly well--determined, it is much less certain how
to connect the tangents on both sides.  From the distribution of
HII-Regions, Georgelin \& Georgelin (1976) sketched a four spiral arm
pattern ranging from about $4\,\KPC$ galactocentric radius to beyond
the solar radius. Although their original pattern has been slightly
modified by later work, the principal result of a four--armed spiral
structure has mostly been endorsed (see the review in Vall\'ee 1995).

For illustration, Fig.~\ref{lvtracers} also shows the traces in the
\lvplot\ of the spiral arms in our standard no--halo model (see
Section 4). This model has two pairs of arms emanating from the end of
the COBE bar, and a four--armed spiral pattern outside the bar's
corotation radius. The model matches the observed tangents rather well
and illustrates the ways in which they can be connected.

\begin{table*}
\begin{tabular}{cccccll}
\multicolumn{5}{c}{Inner Galaxy spiral arm tangents in
longitude}&Measurement&Ref.\\
Scutum & Sagittarius & Centaurus & Norma & 3-kpc& & \\
\hline
29         & 50 & -50  & -32  &     &     HI        &  Weaver (1970), 
Burton \& Shane (1970), Henderson (1977) \\
24, 30.5 & 49.5 & -50 & -30 &     & integrated $^{12}$CO &
Cohen \etal\ (1980), Grabelsky \etal\ (1987) \\
25, 32 & 51  &     &     &     & $^{12}$CO clouds  & Dame \etal\ (1986) \\
25, 30      & 49  &     &     &     & warm CO clouds & Solomon \etal\
(1985)\\
24, 30      & 47 & -55 & -28 &     & HII-Regions (H109-$\alpha$) &
Lockman (1979), Downes \etal\ (1980) \\
32         & 46 & -50 & -35 &     & $^{26}$Al   & Chen \etal\ (1996)\\
32      & 48  &-50,-58& -32 & -21 & Radio $408\,\MHZ$ & Beuermann \etal\
(1985)\\
29         &    &     & -28 & -21 & $2.4\,\MYM$ & Hayakawa \etal\ (1981)
\\
26         &    & -47 & -31 & -20 & $60\,\MYM$  & Bloemen \etal\ (1990)
\\
\hline
30   & 49  & -51 & -31 & -21 & adopted mean &\\
\hline
$\sim25$& 54& -44 & -33 & -20 & \multicolumn{2}{l}{$R_c=3.4\kpc$,
$\PHIBAR=20^\circ$, without halo}\\
$\sim30$& 50& -46 & -33 & -20 & \multicolumn{2}{l}{$R_c=3.4\kpc$, 
$\PHIBAR=20^\circ$, with halo $v_0=200\kms$} \\
$\sim29$& 51& -47 & -34 & -22 & \multicolumn{2}{l}{$R_c=3.4\kpc$, 
$\PHIBAR=25^\circ$, with halo $v_0=200\kms$} \\
\hline
\end{tabular}


\caption[]{%
Observed spiral arm tangents compared to model predictions.
}
\label{armstable}
\end{table*}

\subsection{Dense gas in the Galactic Centre}

Dense regions, where the interstellar medium becomes optically thick
for 21-cm line or \twco\ emission, can still be observed using
rotational transition lines of $^{13}$CO, CS, and other rare species. In the
$^{13}$CO emission line, which probes regions with $\sim 40$ times
higher volume densities than the $^{12}$CO line, an asymmetric
parallelogram--like structure between $\sim+1.5\deg$ and
$\sim-1^\circ$ in longitude is visible (Bally \etal 1988). This is
nearly coincident with the peak in the terminal velocity curve and has
been associated with the cusped $x_1$-orbit by Binney \etal (1991);
see also Section 4. In this interpretation, part of the asymmetry is
accounted for by the perspective effects expected for this elongated
orbit with a bar orientation angle $\phibar=20\deg$.

At yet higher densities, the CS line traces massive molecular cloud
complexes, which are presumably orbiting on $x_2$-orbits inside the
bar's Inner Lindblad Resonance (Stark \etal 1991, Binney \etal
1991). These clouds appear to have orbital velocities of $\lta
100\kms$.

\subsection{Tilt and asymmetry}

We should finally mention two observational facts that are not
addressed in this paper. Firstly, the gaseous disk between the
$x_2$-disk and the 3-kpc-arm is probably tilted out of the Galactic
plane.  Burton \& Liszt (1992) give a tilt angle of about $13 \deg$
for the HI distribution and show that, by combining this with the
effects of a varying vertical scale height, the observed asymmetry in
this region can be explained. Heiligman (1987) finds a smaller tilt
of $\sim 7\deg$ for the parallelogram.

Secondly, the molecular gas disk in the Galactic Center is highly
asymmetric. Three quarters of the \thco\ and CS emission comes from
positive longitudes and a different three quarters comes from material
at positive velocities (Bally \etal 1988). Part of the longitude
asymmetry may be explained as a perspective effect, and part of both
asymmetries is caused by the one--sided distribution of the small
number of giant cloud complexes. Nonetheless it is possible that the
observed asymmetries signify genuine deviations from a triaxially
symmetric potential.

\subsection{Solar radius and velocity -- comparing real and model \lvplot s}

A model calculation results in a velocity field as a function of
position, with the length--scale set by the distance to the Galactic
Center assumed in the deprojection of the COBE bulge (Binney, Gerhard
\& Spergel 1997).  These authors took $R_0=8\kpc$, and throughout this
paper we will use this value in comparing our models to observations.
To convert model velocity fields into \lvplot s as viewed from the
LSR, we scale by a constant factor (this gives the inferred
mass--to--light ratio) and then subtract the line--of--sight component
of the tangential velocity of the LSR, assuming $\VSUN=200\kms$. This
value is in the middle of the range consistent with various
observational data (Sackett 1997), and is also a reasonable value to
use if the Galactic potential near the Sun is slightly elliptical
(Kuijken \& Tremaine 1994). If the model has a constant circular
rotation curve, i.e., if it includes a dark halo, the LSR velocity is
part of the model and is scaled together with the gas velocities.  For
these models the final scaled LSR tangential velocity will be different from
$\VSUN=200\kms$ and will be stated in the text. The radial velocity
of the LSR has been set to zero throughout this paper.

\section{The models}

In this Section, we describe in more detail the models that we use to
study the gas flow in the gravitational potential of the Galactic disk
and bulge, as inferred from the COBE/DIRBE NIR luminosity
distribution. In some of these models the gravitational field of a
dark halo component is added. Self--gravity of the gas and spiral arms
are not taken into account until \S\ref{sec-gravitating}. In the
following, we first describe our mass model as derived from the
COBE/DIRBE NIR data (\S3.1), then the resulting gravitational
potential (\S3.2) and closed orbit structure (\S3.3), the assumptions
going into the hydrodynamical model (\S3.4), and finally the main free
parameters in the model (\S3.5).

\subsection{Mass model from COBE NIR luminosity distribution}

The mass distribution in the model is chosen to represent the luminous
mass distribution as closely as possible. From the NIR surface brightness
distribution as observed by the COBE/DIRBE experiment, Spergel \etal
(1996) computed dust--corrected NIR maps of the bulge region using a
three--dimensional dust model. These cleaned maps were deprojected by
the non--parametric Lucy--Richardson algorithm of Binney \& Gerhard
(1996) as described in Binney, Gerhard \& Spergel (1997; BGS). The
resulting three--dimensional NIR luminosity distributions form
the basis of the mass models used in this paper.

The basic assumption that makes the deprojection of BGS work is that
of eight--fold triaxial symmetry, i.e., the luminosity distribution is
assumed to be symmetric with respect to three mutually orthogonal
planes.  For general orientation of these planes, a barred bulge will
project to a surface brightness distribution with a noticeable
asymmetry signal, due to the perspective effects for an observer at
$8\kpc$ distance from the Galactic Center (Blitz \& Spergel
1991). Vice--versa, if the orientation of the three symmetry planes is
fixed, the asymmetry signal in the data can be used to infer the
underlying triaxial density distribution (Binney \& Gerhard
1996). Because of the assumed symmetry, neither spiral structure nor
lopsidedness can be recovered by the eight--fold algorithm. However,
spiral arm features in the NIR luminosity may be visible in the
residual maps, and may appear as symmetrized features in the recovered
density maps.

The orientation of the three orthogonal planes is specified by two
angles. One of these specifies the position of the Sun relative to the
principal plane of the bulge/bar; this angle takes a well--determined
(small) value such that the Sun is approximately $14\pc$ above the
equatorial plane of the inner Galaxy (BGS). The other angle $\PHIBAR$
specifies the orientation of the bar major axis in the equatorial
plane relative to the Sun--Galactic Center line; this angle is not
well--determined by the projected surface brightness distribution.
However, for a fixed value of the assumed $\PHIBAR$, an essentially
unique model for the recovered 3D luminosity distribution results: BGS
demonstrated that their deprojection method converges to essentially
the same solution for different initial luminosity distributions used
to start the iterations (see also Bissantz \etal 1997).  As judged
from the surface brightness residuals, the residual asymmetry map, and
the constraint that the bar axial ratio should be $<1$, admissable
values for the bar inclination angle $\PHIBAR$ are in the range of
$15^\circ$ to $35^\circ$. We will thus investigate gas flow models
with $\PHIBAR$ in this range.

For the favoured (BGS) $\PHIBAR=20\deg$, the deprojected luminosity
distribution shows an elongated bulge/bar with axis ratios 10:6:4 and
semi--major axis $\sim 2\kpc$, surrounded by an elliptical disk that
extends to $\sim 3.5\kpc$ on the major axis and $\sim 2\kpc$ on the
minor axis.  Outside the bar, the deprojected NIR luminosity
distribution shows a maximum in the emissivity $\sim 3\kpc$ down the
minor axis, which appears to correspond to the ring--like structure
discussed by Kent, Dame \& Fazio (1991). The nature of this feature is
not well understood.  Possible contributions might come from stars on
orbits around the Lagrange points (this appears unlikely in view of
the results of \S3.3 and \S4.1 below), or from stars on $x_1$-orbits
outside corotation or on the diamond shaped $1:4$ resonant orbits
discussed by Athanassoula (1992a) (however, the feature is very
strong). The most likely interpretation in our view, based on \S4
below, is that this feature is due to incorrectly deprojected
(symmetrized) strong spiral arms. If this interpretation is correct,
then by including these features in our mass model we automatically
have a first approximation for the contribution of the Galactic spiral
arms to the gravitational field of the Galaxy.

In the following, we will model the distribution of luminous mass in
the inner Galaxy by using the deprojected DIRBE L--band luminosity
distributions for $15^\circ< \PHIBAR < 35^\circ$ and assuming a
constant L--band mass--to--light ratio $M/L_L$.  The assumption of
constant $M/L_L$ may not be entirely correct if supergiant stars
contribute to the NIR luminosity in star forming regions in the disk
(Rhoads 1998); this issue will be investigated and discussed further
in \S4.1. To obtain a mass model for the entire Galaxy, we
must extend the luminous mass distribution of BGS, by
adding a central cusp and a model for the outer disk, and (in some
cases) add a dark halo to the resulting gravitational potential.

\subsubsection{Cusp}

The density distribution of stars near the Galactic Center can be
modelled as a power law $r^{-p}$. From star counts in the K--band the
exponent $p\simeq 2.2\pm0.2$ for K=6-8 mag stars (Catchpole, Whitelock
\& Glass 1990). The distribution of OH/IR stars near the center gives
$p\simeq 2.0\pm 0.2$ (Lindqvist, Habing \& Winnberg 1992). Using
radial velocities of the OH/IR stars and the assumption of isotropy,
Lindqvist \etal determined the mass distribution inside $\sim
100\pc$. The corresponding mass density profile has $p\simeq 1.5$
between $\sim 20\pc$ and $\sim 100 \pc$ and steepens inside $\sim
20\pc$.  The overall slope is approximately that originally found by
Becklin \& Neugebauer (1968, $p\simeq 1.8$).

In the density model obtained from the DIRBE NIR data, this central
cusp is not recovered because of the limited resolution and grid
spacing ($1.5^\circ$) in the dust corrected maps of Spergel \etal
(1996).  The cusp slope of the deprojected model just outside
$1.5^\circ$ moreover depends on that in the initial model used to
start the Lucy algorithm. To ensure that our final density model
includes a central cusp similar to the observed one, we have therefore
adopted the following procedure. For the initial model used in the
deprojection, we have chosen a cusp slope of $p=1.8$, in the middle of
the range found from star counts and mass modelling. This gives a
power law slope of $p=1.75$ in the final deprojected density model at
around $400\pc$. We have then expanded the deprojected density in
multipoles $\rho_{lm}(r)$ and have fitted power laws to all
$\rho_{l0}$ in the radial range $350-500\pc$.  Inside $350\pc$ these
density multipoles were then replaced by the fitted power laws,
extrapolating the density inwards.  The $m\not=0$ terms were not
changed; they decay to zero at the origin. By this modification the
mass inside $350\pc$ is approximately doubled. The implied change in
mass is small compared to the total mass of the bulge and is 
absorbed in a slightly different mass-to-light ratio when scaling the
model to the observed terminal velocity curve.

\begin{figure}
\getfig{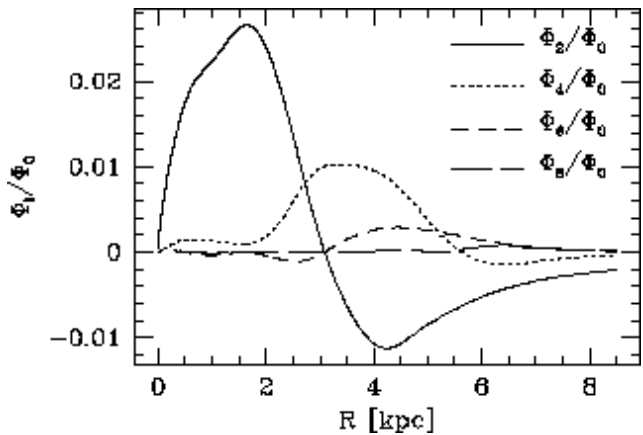}{6cm}
\caption[]{%
Contribution of various planar multipoles to the potential of the
standard $\PHIBAR=20\deg$ bar model. 
}
\label{multipoles}
\end{figure}

\begin{figure}
\getfig{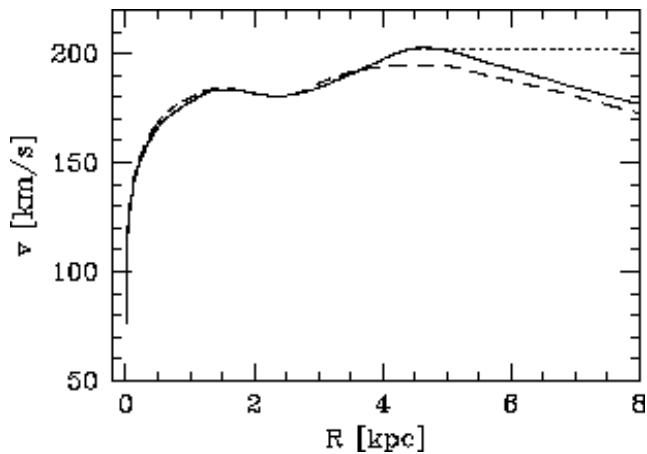}{6cm}
\caption[]{%
Rotation curve in the standard $\PHIBAR=20\deg$ model (solid line), in
the model with substracted ring (dashed), and a model with constant
outer rotation curve (dotted).  The rotation curves are given for a
scaling constant of $\xi=1.075$.
}    
\label{rotationcurve}
\end{figure}

\subsubsection{Outer disk}

The deprojected luminosity model of BGS gives the density in the range
$0<x,y<5\,\KPC$ and $0<z<1.4\,\KPC$. We thus
need to use a parametric model for the mass density of the Galactic
disk outside $R=5\kpc$. In this region the NIR emission is approximated
by the analytic double--exponential disk model given by BGS.  The
least--squares fit parameters are $R_d=2.5\kpc$ for the radial
scale--length, and $z_0=210\pc$ and $z_1=42\pc$ for the two vertical
scale--heights.  These parameters are very similar to those obtained
by Kent, Dame \& Fazio (1991) from their SPACELAB data.
To convert this model for the outer disk luminosity into a mass
distribution, we have assumed that the disk has the same $M/L_L$ as
the bulge, because we cannot distinguish between the bulge and disk
contributions to the NIR emission in the deprojected model for the
{\sl inner} Galaxy.

\subsection{Gravitational potential}

From the density model, we can compute the expansion of the potential
in multipole components $\Phi_{lm}(r)$ and hence the decomposition
\eq{ \Phi(r,\varphi) = \Phi_0(r) +\Phi_2(r,\varphi) \cos(2\varphi)
+\Phi_4(r,\varphi) \cos(4\varphi) } in monopole $\Phi_0$, quadrupole
$\Phi_2$, and octupole $\Phi_4$ terms.  Higher order terms do not
contribute enough to the forces to change the gas flow significantly
(see Fig.~\ref{multipoles}), and are therefore neglected in the
following.  The advantage of this multipole approximation is that it
is economical in terms of computer time (no numerical derivatives are
needed for the force calculations). The quality of the expansion for
the forces was tested with an FFT solver. Errors due to the truncation
of the series are typically below $5\kms$ in velocity. Because this is
less than the sound speed, such errors will not significantly affect
the gas flow.

As already mentioned, it is necessary to modify the potential near the
center because of the unresolved central cusp in the COBE density
distribution. As described above, we have replaced the multipole
components of the density by power law fits inwards of $r=350\pc$
before computing the corresponding multipole expansion of the
potential.  Since the higher order multipoles $\rho_{l0}$ have smaller
power law exponents than the $\rho_{00}$ term, this implies that the
cusp becomes gradually spherically symmetric at small radii.  We have
chosen this approach because it did not require a specification of the
shape of the central cusp. Note that without including the modified
cusp the gravitational potential would not possess $x_2$ orbits and
therefore the resulting gas flow pattern would be different.
Fig.~\ref{multipoles} shows the contribution of the various multipoles
to the final COBE potential of our standard model with
$\varphi=20\deg$. The rotation curve obtained from this potential is
shown in Fig~\ref{rotationcurve}.

\subsubsection{Dark halo}

If the Galactic disk and bulge are maximal, i.e., if they have the
maximal mass--to--light ratio compatible with the terminal velocities
measured in the inner Galaxy, then we do not require a significant
dark halo component in the bar region. This may be close to the
true situation because even with this maximal $M/L_L$ the mass
in the disk and bulge fail to explain the high optical depth
in the bulge microlensing data (Udalski \etal 1994, Alcock \etal 1997)
by a factor $\gta 2$ (Bissantz \etal 1997). Thus in our modelling of
the bar properties we have not included a dark halo component.

However, for the spiral arms found outside corotation of the bar, the
dark halo is likely to have some effect. Since we only study the gas
flow in the galactic plane, the force from the dark halo is easy to
include without reference to its detailed density distribution. We
simply change the monopole moment in the potential directly such that
the asymptotic rotation curve becomes flat with a specified circular
velocity. The rotation curve for our flat rotation model is also shown
in Fig~\ref{rotationcurve}. In this model, the halo contribution to
the radial force at the solar circle is $\simeq 23\%$.

\subsection{Effective potential, orbits, and resonance diagram}

\begin{figure*}
\ifpsfiles\centerline{\psfig{figure=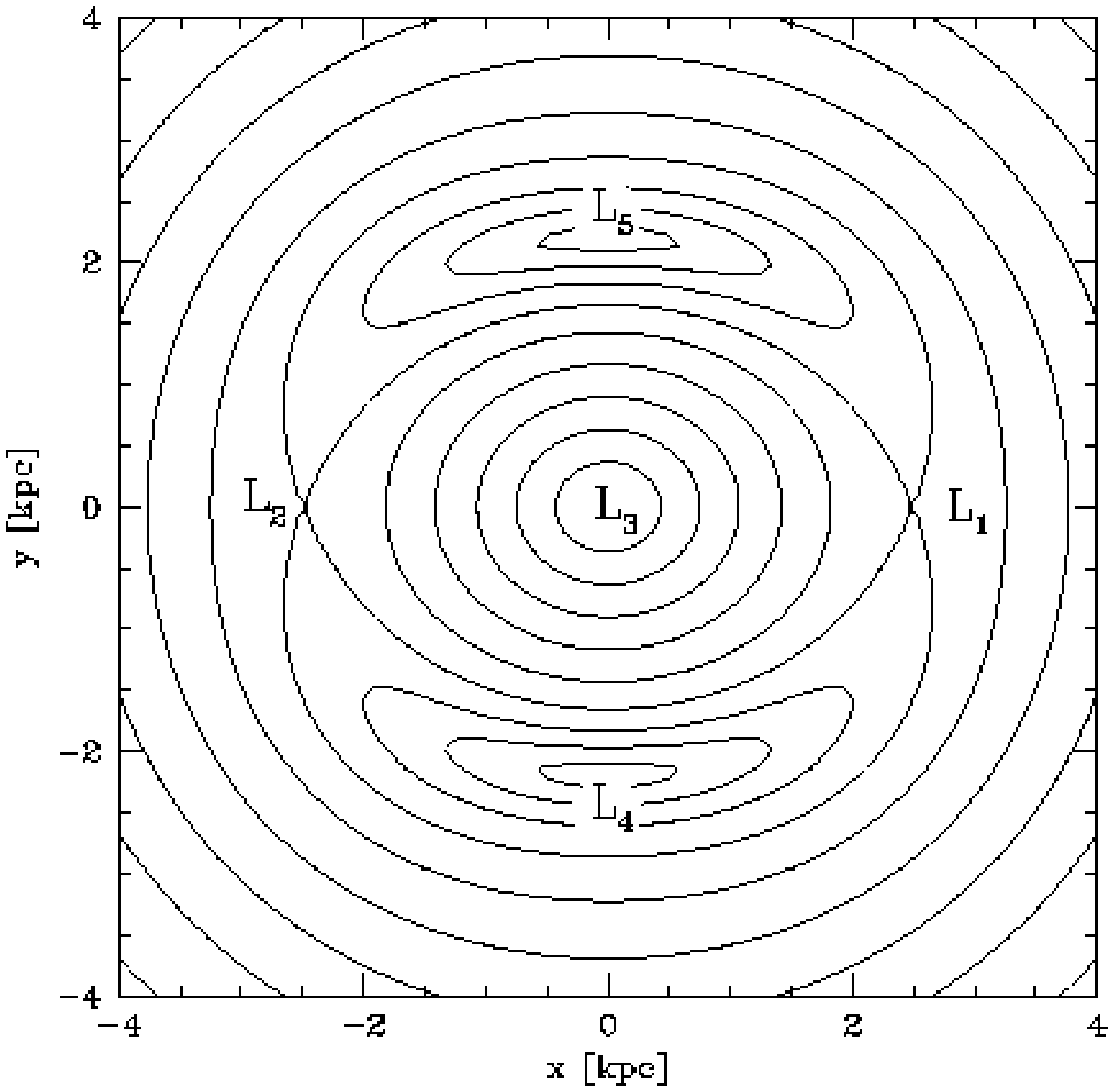,width=8.2cm}
	        \quad \psfig{figure=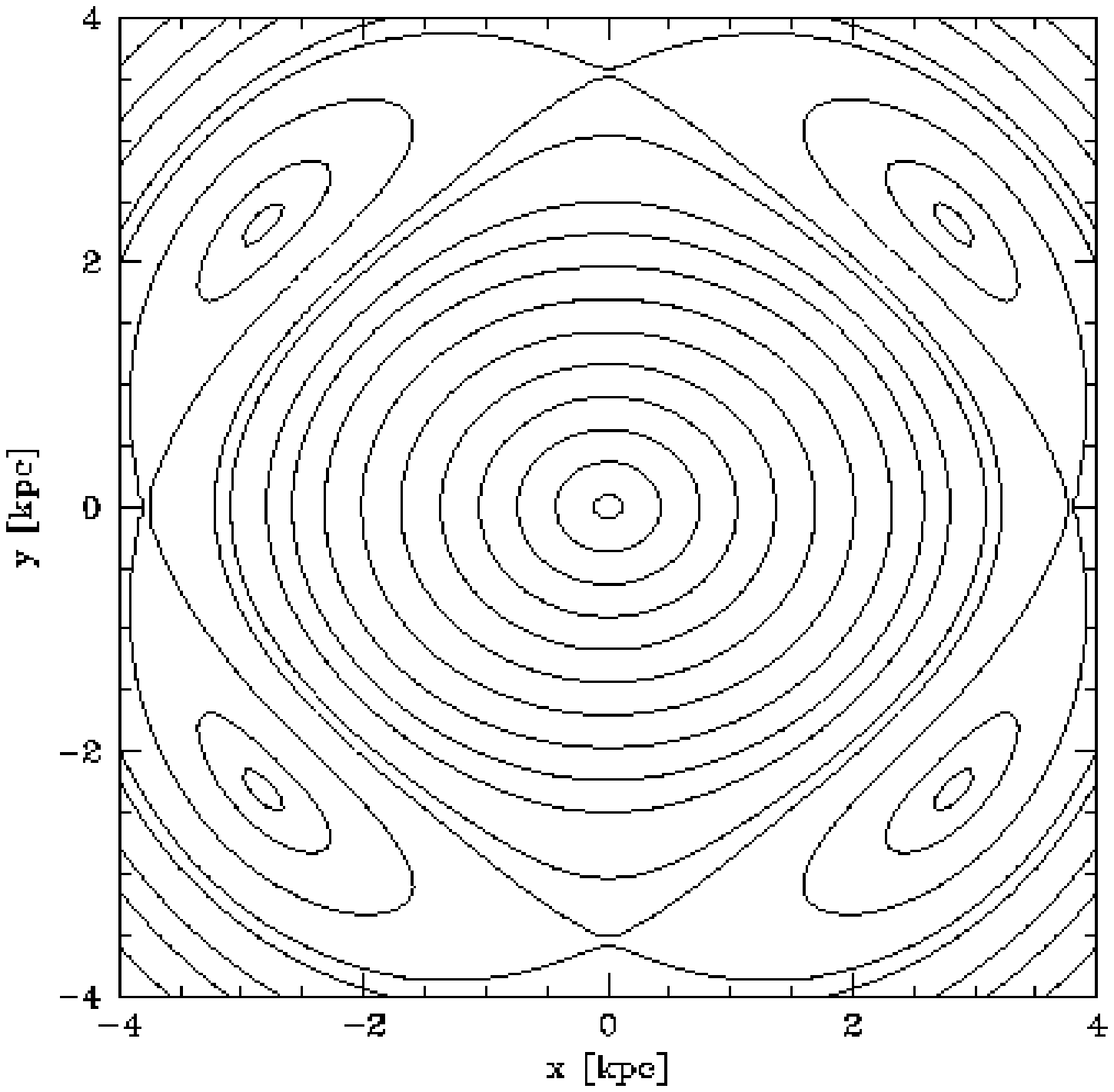,width=8.2cm}}
\else\vskip6cm\fi
\caption[]{%
Left: Effective potential in the standard $\PHIBAR=20\deg$ bar model
for $\OmegaP=80\,\KMSKPC$, showing the usual four Lagrangian points in
the corotation region.  Right: For $\OmegaP=55\,\KMSKPC$. Because the
mass peaks in the disk $\sim 3\kpc$ down the bar's minor axis now
contribute significantly to the potential near the increased
corotation radius, there are eight Lagrangian points near corotation
for this pattern speed.
}
\label{phieffective}
\end{figure*}

\begin{figure*}
\ifpsfiles\centerline{\psfig{figure=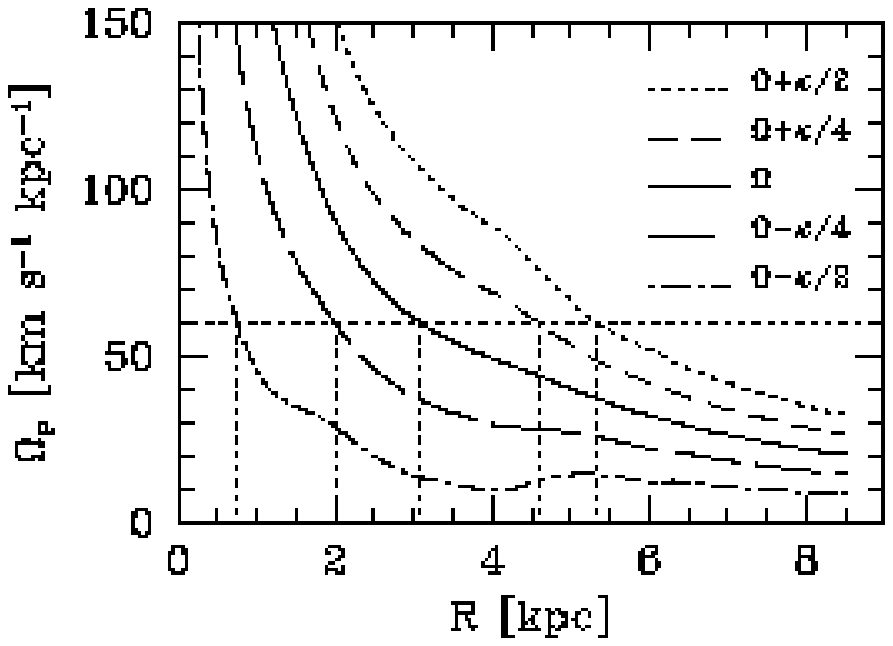,width=8.2cm}
	        \quad \psfig{figure=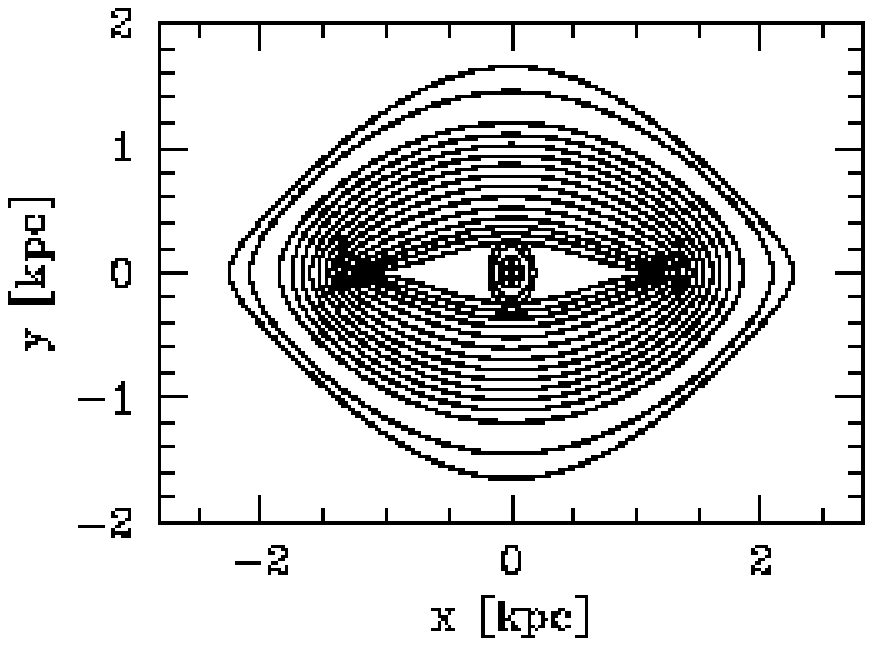,width=8.2cm}}
\else\vskip6cm\fi
\caption[]{%
Left: Resonance diagram for the standard $\PHIBAR=20\deg$ bar model.
Right: Some $x_1$ and $x_2$ orbits in this model for a pattern
speed of $60\,\KMSKPC$. The gap between the second and third
orbit from outside shows the location of the $1:4$ resonance.
}
\label{resdiagram}
\end{figure*}

The constructed galaxy models have some special properties, due to the
mass peaks in the disk $\sim 3\kpc$ down the minor axis of the bar. In
the more common barred galaxy models, the effective potential \eq{
\Phieff = \Phi - {1\over2} \Omega_P^2 R^2 } in the rotating bar frame
contains the usual four Lagrange points around corotation and a fifth
Lagrange point in the centre.  In our case, this is true only for
larger pattern speeds $\Omega_P$, say $80\,\KMSKPC$, when the
corotation region does not overlap with the region affected by these
(presumably) spiral arm features.  For lower pattern speeds, in
particular for $\Omega_P\simeq 55-60\,\KMSKPC$ which we will find
below to be appropriate for the Milky Way bar, the situation is
different: in this case we obtain four stable and four unstable
Lagrange points around corotation. The four unstable points lie along
the principal axes where normally the four usual Lagrange points are
located, whereas the stable Lagrange points lie between these away
from the axes (see Fig. \ref{phieffective}).  For yet lower pattern
speeds, the number of Lagrange points reduces to four again, but then
the two usual saddle points have changed into maxima and vice versa.

We have not studied the orbital structure in this potential in great
detail. However, some of the orbits we have found are shown in
Fig. \ref{resdiagram}, demonstrating the existence of $x_1$, $x_2$,
and resonant 1:4 orbits also in this case when there are eight
Lagrange points near corotation. This is presumably due to the fact
that these orbits do not probe the potential near corotation.  For the
orbit nomenclature used here see Contopoulos \& Papayannopoulos
(1980).

With appropriate scaling, the envelope of the $x_1$ orbits in our
model follows the observed terminal curve in the longitude--velocity
diagram, and the peak in the curve corresponds approximately to the
cusped $x_1$ orbit, as in the model of Binney \etal (1991). This will
be discussed further in \S4.

\subsection{Hydrodynamical Models}

For the hydrodynamical models we have used the two--dimensional
smoothed particles hydrodynamics (SPH) code described in Englmaier \&
Gerhard (1997). The gas flow is followed in the gravitational
potential of the model galaxy as given by the multipole expansion
described in \S3.2, in a frame rotating with a fixed pattern speed
$\OmegaP$. In some later simulations we have included the self-gravity
of the spiral arms represented by the gas flow (see
\S\ref{sec-gravitating} below).

All models assume point symmetry with respect to the centre. This
effectively doubles the number of particles and leads to a factor of
$\sqrt{2}$ improvement in linear resolution.  We have checked that
models without this symmetry give the same results, as would be
expected because the background potential dominates the dynamics.

\label{sec-scaling-def}

The hydrodynamic code solves Euler's equation for an isothermal gas
with an effective sound speed $c_s$:
\eq{
  {\partial{\mathbf v}\over\partial t} + ({\mathbf v}\cdot\nabla){\mathbf v} =
	-\csnd^2{\nabla \rho\over\rho} - \nabla\Phi.
}
This is based on the results of Cowie (1980) who showed that a crude
approximation to the ISM dynamics is given by an isothermal single
fluid description in which, however, the isothermal sound speed is not
the thermal sound speed, but an effective sound speed representing the
\rms\ random velocity of the cloud ensemble.

Using the SPH method to solve Eulers equation has the advantage to
allow for a spatially adaptive resolution length. The smoothing length
$h$, which can be thought of denoting the particle size, is adjusted
by demanding an approximately constant number of particles overlapping
a given particle. The SPH scheme approximates the fluid quantities by
averaging over neighboring particles and, in order to resolve shocks,
includes an artificial viscosity. This can be understood as an
additional viscous pressure term which allows the pre-shock region to
communicate with the post-shock region, i.e., to transfer momentum. We
have used the standard SPH viscosity (Monaghan \& Gingold 1983) with
standard parameters $\alpha=1$ and $\beta=2$. This SPH method was
tested for barred galaxy applications by verifying that the properties
of shocks forming in such models agree with those found by
Athanassoula (1992b) with a grid-based method. See Steinmetz \&
M\"uller (1993) and Englmaier \& Gerhard (1997) for further details.

In the low resolution calculations described below we have generally
used 20000 SPH particles and have taken a constant initial surface
density inside $7\kpc$ galactocentric radius.  With the assumptions
that the gas flow is two--dimensional and point--symmetric, these
parameters give an initial particle separation in the Galactic plane
of $62\pc$. High resolution calculations include up to 100,000 SPH
particles and may cover a larger range in galactocentric radius to
investigate the effects of the outer boundary.

We have experimented with two methods for the initial setup of the gas
distribution.  Method A starts the gas on circular orbits in the
axisymmetric part $\Phi_0$ of the potential.  Then the
non-axisymmetric part of $\Phi$ is gradually introduced within
typically one half rotation of the bar. Method B places the gas on
$x_1$-orbits outside and on $x_2$-orbits inside the cusped
$x_1$-orbit. The latter method leads to a more quiet start than the
former, since the gas configuration is already closer to the final
equilibrium. Most models shown in this paper have been set up with
Method A. One model was created with a combination of both methods to
improve resolution around the cusped orbit (see \S~\ref{sec-termcurve}).

Different models are usually compared at an evolutionary age of
$0.3\gyr$, when the gas flow has become approximately quasi-stationary
(see \S4.2). This corresponds to just under three particle rotation
periods at a radius of $3\kpc$. The turn-on time of the bar is $\sim
0.04\gyr$ and is included in the quoted evolution age.

Since the mass--to--light ratio of the model is not known a priori,
all velocities in the model are known only up to a uniform scaling
constant $\xi$. This implies that also the final sound speed is scaled
from the value used in the numerical calculation: If we know the
solution $\mathbf v(\mathbf r)$ in one potential $\Phi$ with galaxy
mass $M$, and sound speed $\csnd$, then we also know the scaled
solution $\xi\mathbf v(\mathbf r)$ for the potential $\xi^2\Phi$,
galaxy mass $\xi^2M$, and sound speed $\xi\csnd$.  For the gas we thus
effectively assume an isothermal equation of state with sound speed
$\csnd=\xi\,10\,\KMS$.  Note that, because $\csnd$ is a local physical
quantity, our model can not simply be scaled down to a dwarf size
galaxy, because then the resulting sound speed would be too small and
this matters because the gas flow pattern depends on this parameter
(Englmaier \& Gerhard 1997).

Below, we fix the scaling constant $\xi$ by fitting the terminal
velocity curve of the model to the observed terminal velocity curve.
To do this, we have to simultaneously assume a value for the local
standard of rest (LSR) circular motion. These two parameters
compensate to some extent, but we generally find that the fit to the
terminal curve is more sensitive to the assumed LSR motion than
to the value of the scaling constant. In most of the rest of the paper
we have therefore fixed the LSR velocity to $200\kms$.
We work in units of $\kpc$, $\gyr$, and $\msun$. If the deprojected
COBE density distribution is assumed to be in units of
$3\,10^8\msun\kpc^{-3}$, $\xi$ is found to have typical values of
1.075 to 1.12.

\subsection{Summary of model parameters and discussion of assumptions}

Our models have a small number of free parameters; these and the
subset which are varied in this paper are listed here. This Section
also contains a brief summary and discussion of the main assumptions.

\subsubsection{Bar parameters}

$\bullet$ Probably the most important parameter is the bar's
corotation radius $R_c$ or pattern speed $\OmegaP$, which will set the
location of the resonance radii and spiral arms and shocks in the gas
flow.

\noindent
$\bullet$ The second important bar parameter is its orientation
angle $\varphi$ with respect to the Sun--Galactic Center line,
which affects the appearance of the gas flow as viewed from
the Sun.

The shape and radial density distribution in the bar region are
constrained by the observed NIR light distribution. However, their
detailed form is dependent on the assumption of 8-fold symmetry, which
is likely to be a good assumption in the central bulge region, but
might be too strong in the outer bar regions, where a possible spiral
density wave might affect the dynamics.  It is also possible that an
overall $m=1$ perturbation is needed to explain the observed
asymmetries, such as in the distribution of giant cloud complexes in
the Galactic center, or the fact that the 3-kpc-arm appears to be much
stronger than its counterarm.  Nevertheless, it is important to find
out how far we can go without these asymmetries. In any case, the
bar should have the strongest impact on the dynamics.

\subsubsection{Mass model}

$\bullet$ The only additional parameter in the luminous mass model
is the scaling constant $\xi$ which relates NIR luminosity
and mass. For each pair of values of the previous two parameters
and at fixed LSR rotation velocity this is determined from
the Galactic terminal velocity curve, assuming that this is
dominated by the luminous mass in the central few $\kpc$.

This contains the additional assumption that all components have the
same constant NIR mass-to-light ratio.  This appears to be a
reasonably good assumption on the basis of the fact that optical--NIR
colours of bulges and disks in external galaxies are very similar
(Peletier \& Balcells 1996). It is unlikely to be strictly correct,
however, because the bulge and disk stars will not all have formed at
the same time. To relax this assumption requires additional
assumptions about the distinction between disk and bulge stellar
luminosity.  This is presently impractical.

\noindent
$\bullet$ Depending on the LSR rotational velocity, a dark halo is
required beyond $R\simeq 5\kpc$. Thus we need to specify the
asymptotic circular velocity of the halo. Here we consider only
two cases, one without halo, the other with asymptotic halo velocity
of $v_0=200\kms$. This is in the middle of the observed range of
$180-220\kms$ (Sackett 1997).

\subsubsection{LSR motion and position}

\noindent$\bullet$
We assume throughout this paper that the distance of the LSR to the galactic
centre is $R_0=8\,\KPC$ (see the review by Sackett 1997, but
also the recent study by Olling \& Merrifield 1998, who argue for a
somewhat smaller $R_0$).

To compare model velocity fields with observations, we have to know
not only the position of the Sun but also its motion. The peculiar
motion of the Sun relative to the LSR is often already corrected for
in the published data. 

\noindent$\bullet$
The remaining free parameter is the LSR rotational velocity
around the galactic center, which lies in the range between
$180-220\kms$ (Sackett 1997). We will again use $\VSUN=200\,\KMS$,
consistent with the above.

\subsubsection{Gas model}

We use a crude approximation to the ISM dynamics, that of an isothermal
single fluid (Cowie 1980). The effective sound speed $c_s$ is the
cloud--cloud velocity dispersion; this varies from $\sim 6\kms$ in the
solar neighbourhood to $\sim 25\kms$ in the Galactic Center gas
disk. We have considered models with globally constant value of $c_s$
between $5-30\kms$ and have not found any interesting effects. Only at the
largest values do the spiral arm shocks become very weak.

Consistent with the assumption of eight--fold symmetry for the mass
distribution, we have assumed the gas flow to be point symmetric with
respect to the origin.  For gas flows in the eight--fold symmetric
potential and without self-gravity there are no significant
differences to the case when the gas model is run without symmetry
constraint.

\section{Results}

\subsection{Gas flow morphology implied by the COBE
luminosity/mass distribution}

\label{sec-morph}

We begin by describing the morphology of gas flows in the
COBE--constrained potentials.  For our starting model we take the
deprojected eight--fold symmetric luminosity distribution obtained
from the cleaned COBE L-band data, for a bar angle $\PHIBAR=20\deg$ as
favoured by BGS. This model, with constant mass--to--light ratio and
no additional dark halo, will be referred to as the (standard)
$\PHIBAR=20\deg$ COBE bar.

In our first simulation this bar model is assumed to rotate at a
constant pattern speed $\Omega_p$ such that corotation is at a
galactocentric radius of approximately $3.1\kpc$.  With the value for
the mass--to--light ratio $\Upsilon_L$ as determined by Bissantz \etal
(1997) from fitting the observed terminal velocities, this gives
$\Omega_p=60\kms/\kpc$.

For the gas model we take a constant initial surface density on
circular orbits, represented by 20000 SPH particles, and an effective
isothermal sound speed of $c_s=10\kms$. The gas is relaxed in the
bar potential as described in \S3.4 (Method A) and the initial
particle separation in the Galactic plane is $62\pc$.


\begin{figure}
\ifpsfiles
\centerline{\psfig{figure=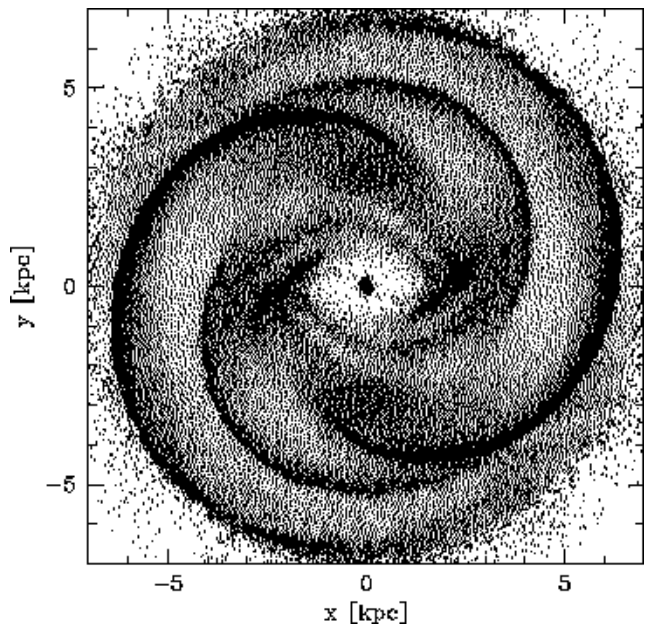,width=\hsize}}
\else\vskip6cm\fi
\caption[]{%
Gas flow in the $\PHIBAR=20\deg$ COBE bar with corotation at
$R_c\simeq 3.1\kpc$. In this and subsequent figures, the long axis
of the bar lies
along the $x$-axis, and the location of the Sun is at $x=-7.5\kpc$,
$y=-2.7\kpc$, $20\deg$ away from the bar's major axis.  The simulation
has $N=20000$ SPH particles with point symmetry built in. The initial
gas disk extends to $R_{\rm max}=7\kpc$.  The multipole expansion of
the potential includes all significant terms (up to $l=6$, $m=4$).
}
\label{phi-om60-phi20-lm64}
\end{figure}

\begin{figure}
\ifpsfiles
  \centerline{\psfig{figure=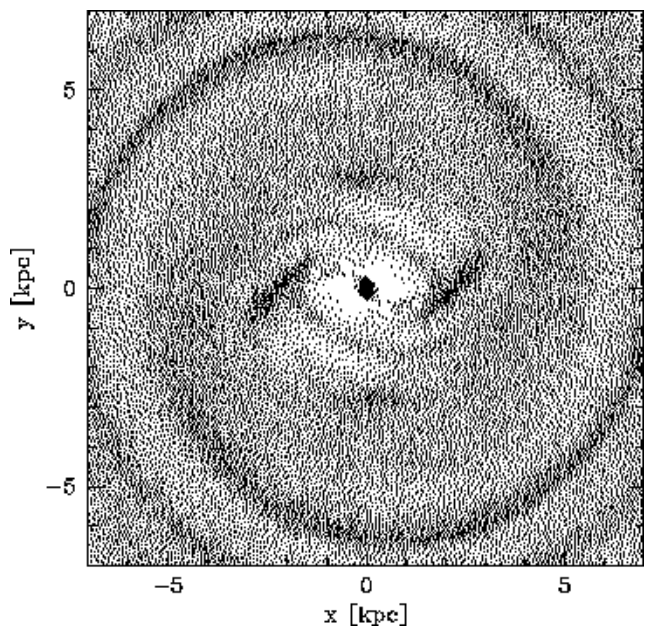,width=\hsize}}
\else\vskip6cm\fi
\caption[]{%
Gas flow in the same model as in Fig.~\ref{phi-om60-phi20-lm64}, but
with the density multipoles with $m\not=0$ set to zero outside
$3\kpc$. Since the $l=0$ terms are unchanged, the circular rotation
curve remains the same.  The gas model extends to $10\kpc$ and has
20'000 particles.
}
\label{phi-om60-phi20-lm62}
\end{figure}

Fig.~\ref{phi-om60-phi20-lm64} shows the morphology of the gas flow in
this model.  Inside corotation ($R_c=3.1\kpc$) two arms arise near
each end of the bar. The dust lane shocks further in are barely
resolved in this model. The structure into the corotation region is
complicated. Outside corotation four spiral arms are seen. This
four--armed structure is characteristic of all gas flow models that we
have computed in the unmodified COBE potentials.  A four--armed spiral
pattern is considered by many papers the most likely interpretation of
the observational material regarding the various spiral arm tracers
and the five main spiral arm tangent points inside the solar circle
(cf.\ Figs.~\ref{obslvplot}, \ref{lvtracers}, Table~\ref{armstable}),
see the review and references in Vall\'ee (1995). 

Two of the four arms emanate approximately near the major axis of the
bar potential, two originate from near the minor axis.  The additional
pair of arms compared to more standard configurations is caused by the
octupole term in the potential; in a model where this term is removed,
the resulting gas flow has only two arms outside corotation.

Fig.~\ref{phi-om60-phi20-lm62} shows the gas flow in a model in which
the density multipoles with $m\not=0$ were set to zero outside $3\kpc$
before computing the potential. This modification leaves the circular
rotation curve of the model unchanged. All structure in the resulting
gas flow is now driven by the rotating bar inside corotation, whose
quadrupole moment outside $R_c$ is weak.  The figure shows that,
correspondingly, only two weak spiral arms now form in the disk
outside corotation.  In the \lvplot, these appear as tangents at
longitudes $l\simeq -50\deg$ and $l \simeq 50\deg$.  However, there
are no arms in this model which would show along the tangent
directions $l=\pm30\deg$; at best there are slight density
enhancements in these parts of the disk. However, the abundance of
warm CO clouds found near $l=25-30\deg$ by Solomon, Sanders \& Rivolo
(1985; cf.\ Fig.~\ref{lvtracers}) indicates that a spiral arm shock
must be present in this region. Thus the model underlying
Fig.~\ref{phi-om60-phi20-lm62}, in which all structure in the gas disk
outside $3\kpc$ is driven by only the rotating bar in the inner
Galaxy, cannot be correct.

Both the quadrupole and octupole terms of the potential outside $\sim
3\kpc$ are dominated by the strong luminosity--mass peaks about
$3\kpc$ down the minor axis of the COBE bar.  From comparing
Figs.~\ref{phi-om60-phi20-lm64} and \ref{phi-om60-phi20-lm62} we thus
conclude that, in order to generate a spiral arm pattern in the range
$R=3-8\kpc$ in agreement with observations these peaks in the NIR
luminosity must have significant mass. In other words, the NIR
mass--to--light ratio in this region cannot be much smaller than the
overall value in the bulge and disk. This result is in agreement with
a recent study by Rhoads (1998) who finds that in external galaxies
the {\sl local} contribution of young supergiant stars to the NIR flux
can be of order $\sim 33\%$ but does not dominate the old stellar
population.

The observed luminosity peaks on the minor axis of the deprojected
COBE bar coincide with dense concentrations of gas particles near the
heads of the two strongest spiral arms in our gas model
(Fig.~\ref{phi-om60-phi20-lm64}). Since the gas arms will generally be
accompanied by stellar spiral arms, this suggests that the most likely
interpretation of the minor axis peaks in the BGS model is in terms of
incorrectly deprojected spiral arms.

Spiral arms are generally the sites of the most vigorous star
formation in disk galaxies. Also in the Milky Way, observations of far
infrared emission show that most of the star formation presently
occurs in the molecular ring (Bronfman 1992). Since we have found from
dynamics that even in this region the associated young supergiants do
not dominate the NIR light, this implies that over most of the
Galactic disk the assumption of constant NIR mass--to--light ratio for
the old stars is justified.

\subsection{Time evolution}

\begin{figure*}
\ifpsfiles\hbox to \hsize{\vbox{
		\centerline{\psfig{figure=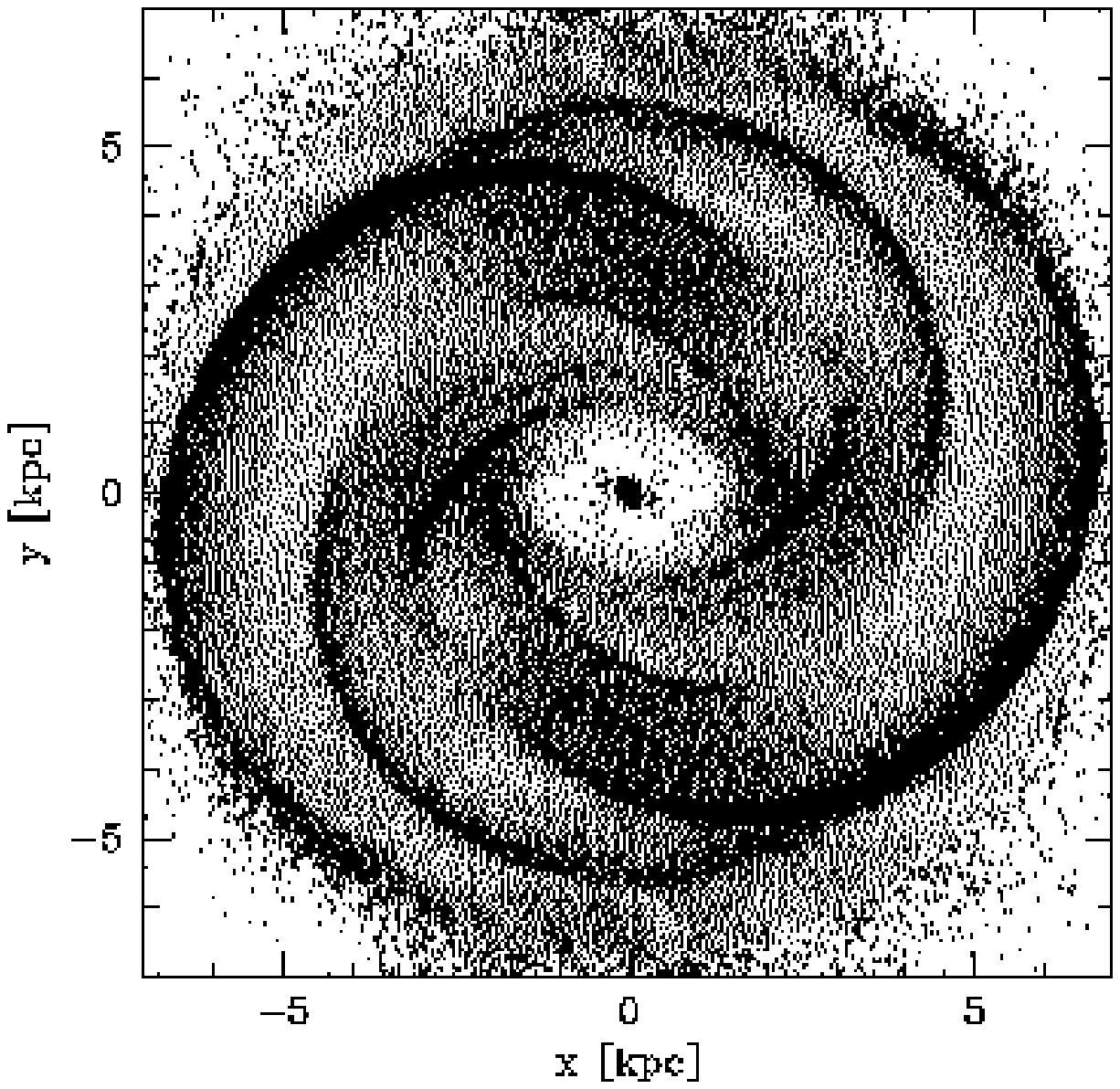,width=8.2cm}
	        \quad \psfig{figure=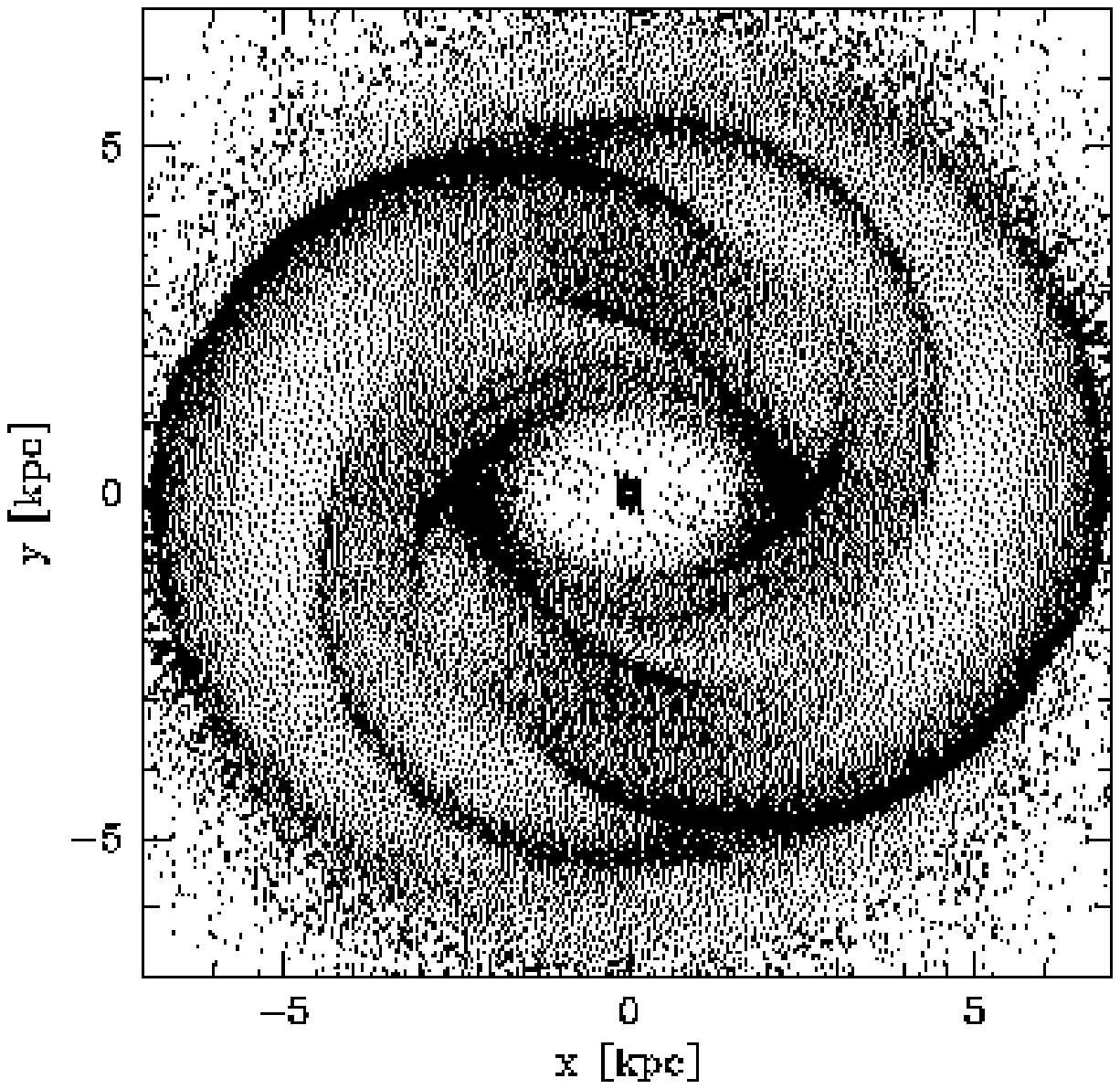,width=8.2cm}}
		\medskip
		\centerline{\psfig{figure=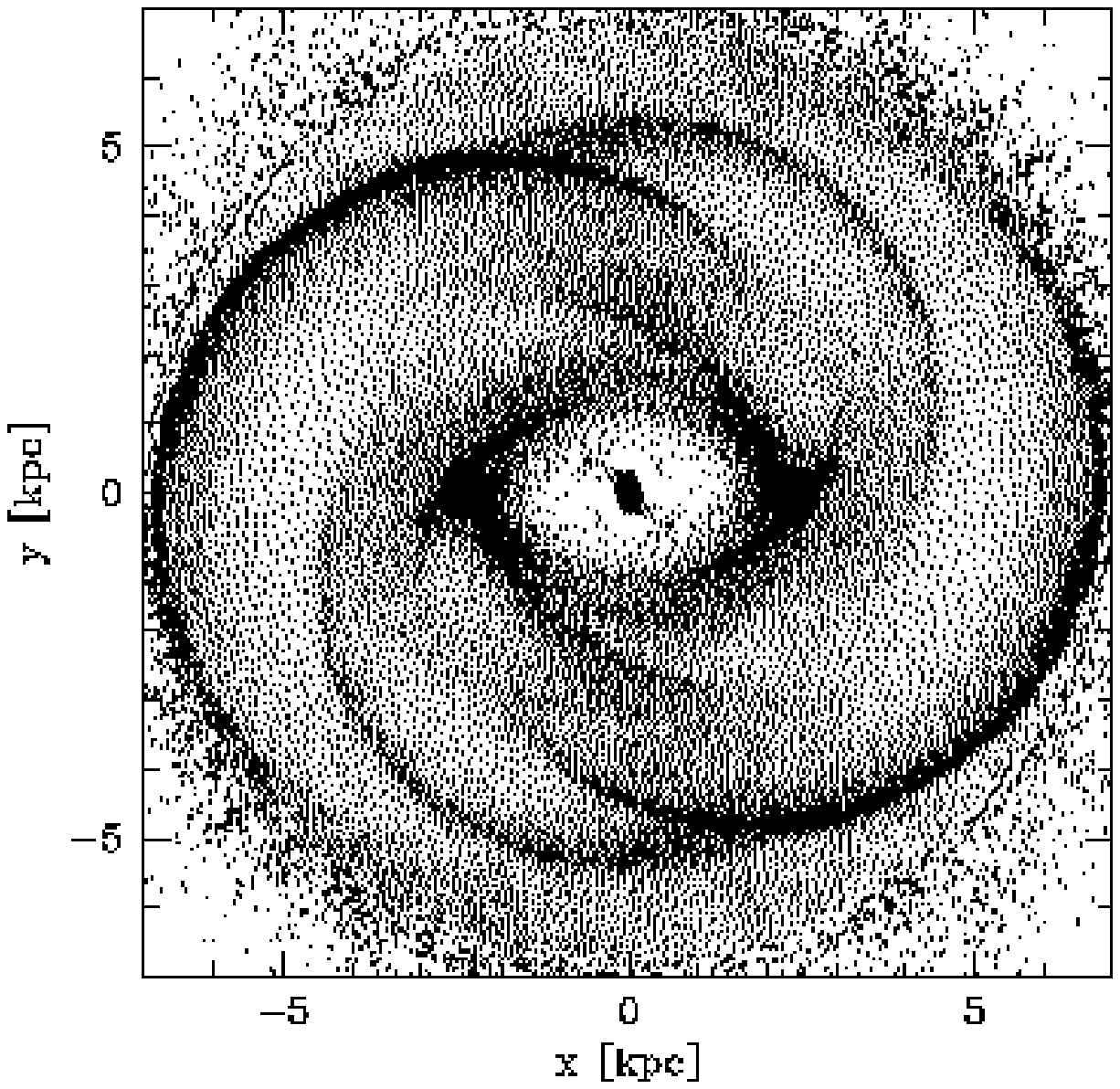,width=8.2cm}
	        \quad \psfig{figure=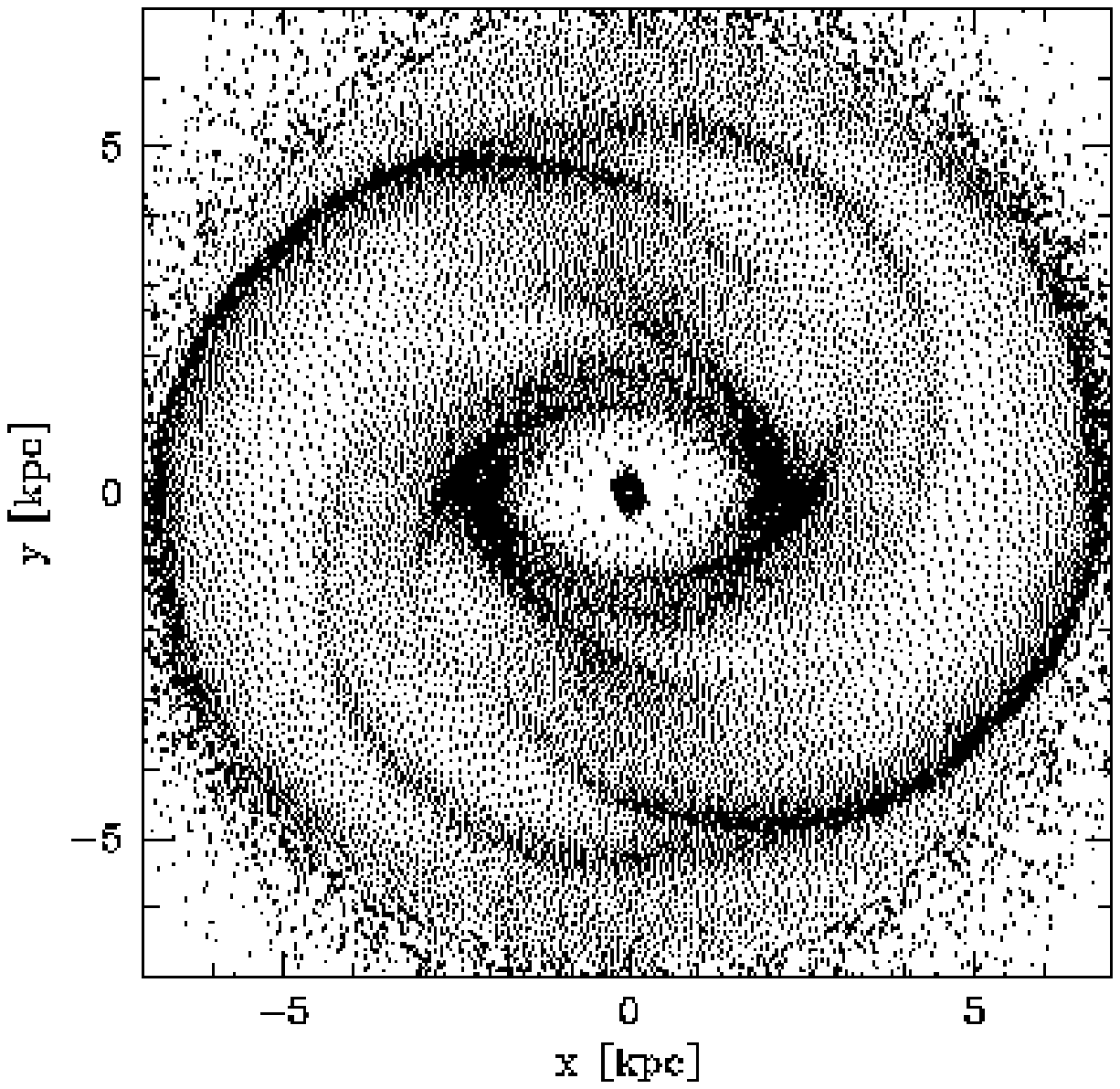,width=8.2cm}}
		} }
\else\vskip12cm\fi
\caption[]{%
Gas particle distribution in the standard $\PHIBAR=20\deg$ COBE bar
potential, for a corotation radius $R_c\simeq 3.4 \kpc$.
The frames show snapshots at $t=0.3\gyr, 1.0\gyr, 2.0\gyr, 3.0\gyr$
(top left, top right, bottom left, bottom right). The bar was 
gradually turned on between $t=0$ and $t=0.04\gyr$. The 
particle rotation period at $3\kpc$ galactocentric radius is
$\simeq 0.1\gyr$. The most significant evolutionary effect is a loss
of resolution by about a factor of two in linear distance between the
first and the last frame, due to substantial mass inflow. This results
in fuzzier spiral arms at the end of the simulation, which appear to
terminate earlier. Also compare Fig.~\ref{lvoftime}.
}
\label{rhooftime}
\end{figure*}

How stationary is the morphological structure in these gas flows?  To
address this question, we show in Fig.~\ref{rhooftime} the time
evolution of a typical model ($\PHIBAR=20\deg$,
$\OmegaP=55\,\KMSKPC$). In this and other simulations the
non-axisymmetric part of the gravitational potential was gradually
turned on within about one half of a bar rotation period ($\simeq 0.04
\gyr$). The gas flow, which is initially on circular orbits, then
takes some time to adjust to the new potential. It reaches a
quasi--stationary pattern by about time $t=0.3\gyr$. This flow is
shown in the top left panel of Fig.~\ref{rhooftime}. After $t=0.3\gyr$
the variations in the gas flow are small: about $5\,\KMS$ in the
velocities.  Also the sharpness of the arms inside corotation varies
slightly.  In this quasi-stationary flow material continuously streams
inwards: Gas parcels that reach a shock dissipate their kinetic energy
perpendicular to the shock. Subsequently they move inwards along the
shock.

As Fig.~\ref{rhooftime} shows, the inward gas inflow causes a slow
evolution without much changing the morpophology of the gas flow.
However, the mass accumulating on the central disk of $x_2$--orbits in
the course of this process is considerable. In fact, to continue the
simulation we have found it necessary to constantly remove particles
from the $x_2$--disk.  In doing this we have simultaneously increased
the particle mass in this region in such a way as to keep the surface
density unchanged. Therefore effectively we have only limited the
resolution in this region from increasing ever further, without
rearranging or changing mass. The gas inflow leads to a loss of
resolution in the outer disk.  From $t=0.3\gyr$ to $t=3\gyr$, the
surface density of particles in outer disk of the model shown in
Fig.~\ref{rhooftime} decreases by about a factor of four, i.e., the
linear resolution by about a factor of two. The spiral arms therefore
become more difficult to see; in particular, the starting points and
end points of some of the arms appear to shift slightly.

The most rapid evolution occurs in the vicinity of the cusped orbit.
Already by time $t=0.3\gyr$, the gas disk near this orbit has been
strongly depleted. Because the shear in the velocity field in the
vicinity of the cusped orbit is very strong, particles that reach the
cusped orbit shock move to the center quickly along the shock ridges
and then fall onto the $x_2$--disk (Englmaier \& Gerhard 1997). In the
low-density region between the cusped orbit and the $x_2$--disk, the
smoothing radius of the SPH particles is large compared to the
velocity gradient scale. It is possible that the resulting large
effective viscosity accelerates the depletion of gas on the cusped
orbit and in its vicinity. A similar effect has been seen by Jenkins
\& Binney (1994) in their sticky particle simulations. See
\S~\ref{sec-termcurve} for an improved model with more resolution in
the cusped orbit region.

\subsection{Model (l,v)-diagrams}

\begin{figure}
\ifpsfiles
	\centerline{\psfig{figure=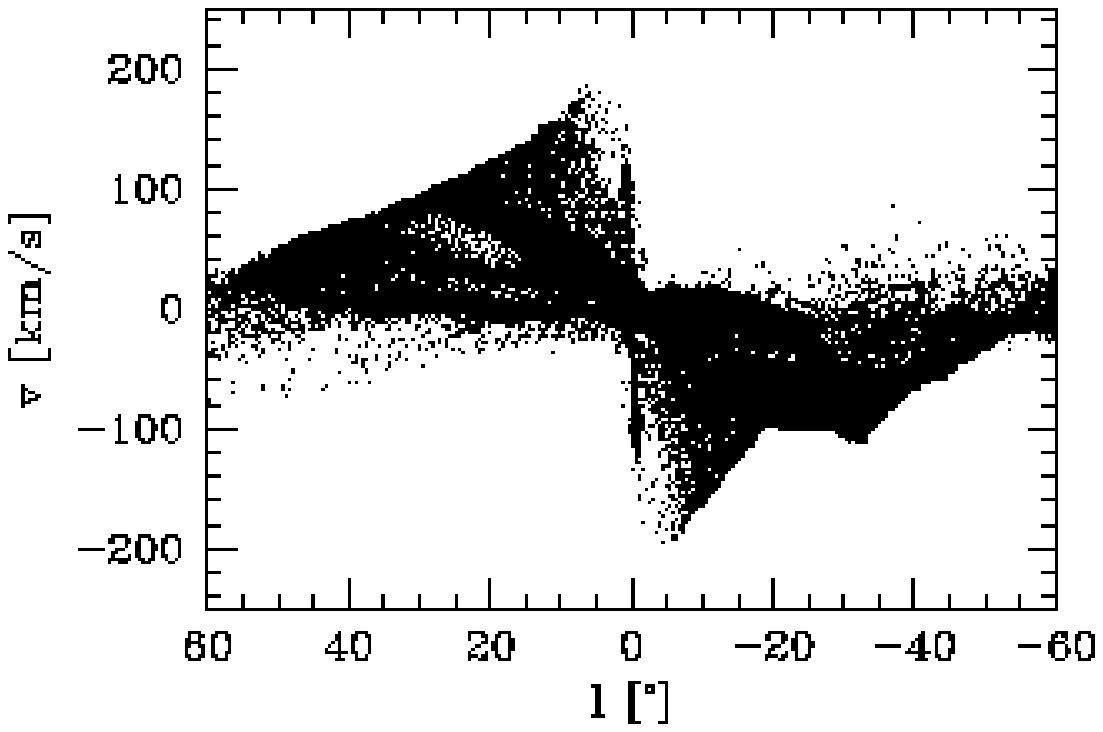,width=8.2cm}}
	\medskip
	\centerline{\psfig{figure=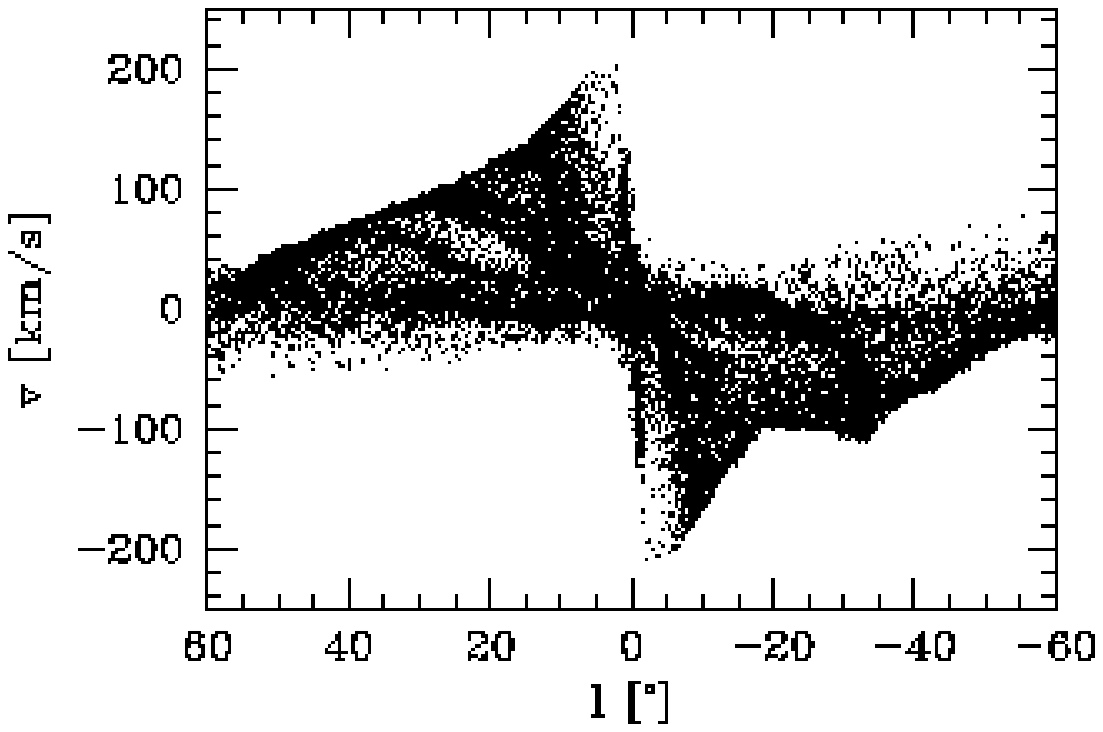,width=8.2cm}}
\else\vskip12cm\fi
\caption[]{%
Longitude--velocity \lvplot s corresponding to the gas particle
distributions in Fig.~\ref{rhooftime} at $t=0.3\gyr$ (top) and
$3\gyr$ (bottom). In constructing these we have assumed $R_0=8\kpc$
and $v_0=200\kms$. The inner disk on $x_2$-orbits,
the terminal velocity curve, and the spiral arm traces are apparent.
}
\label{lvoftime}
\end{figure}

\begin{figure}
\ifpsfiles
\centerline{\psfig{figure=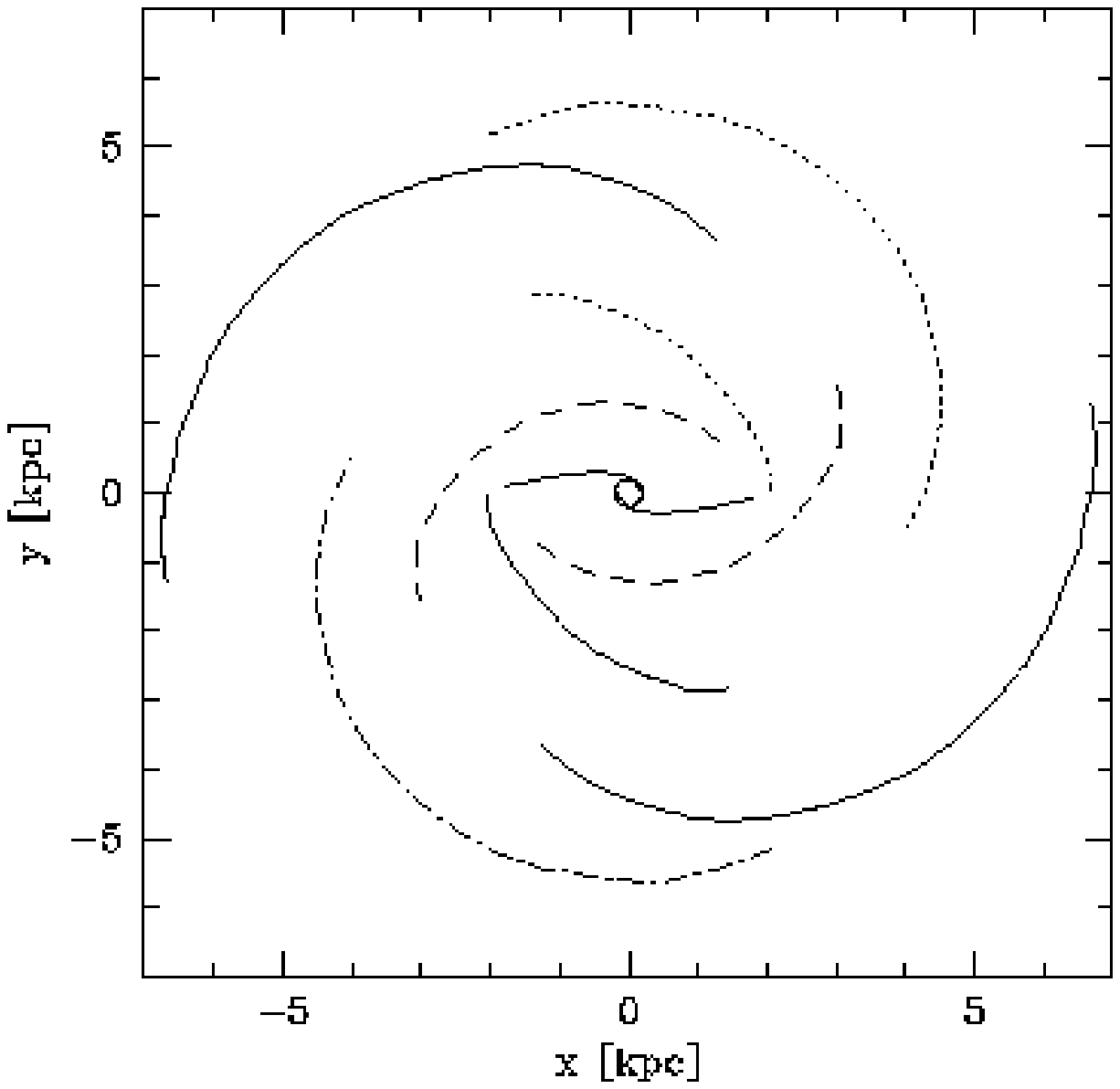,width=\hsize}}
\centerline{\psfig{figure=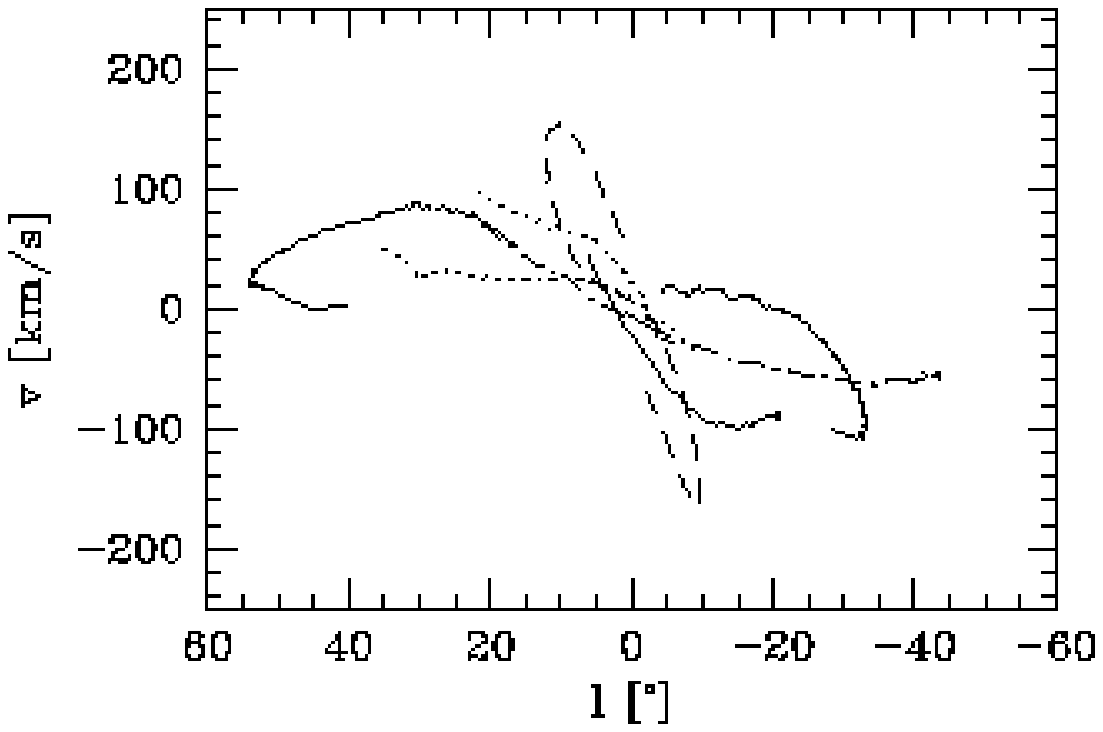,width=\hsize}}
\else\vskip12cm\fi
\caption[]{%
 Top: Schematic representation of the spiral arms in the gas flow model
depicted in the top left frame of Fig.~\ref{rhooftime}.

Bottom: The same spiral arms in the \lvplot.
}
\label{lvexplain}
\end{figure}

The non-axisymmetric structures seen in Figs.~\ref{rhooftime} etc.\
lead to perturbations of the gas flow velocities away from circular
orbit velocities. These can be conveniently displayed in an \lvplot\ 
like those often used for representing Galactic radio observations.
In fact, to constrain the Galactic spiral arm morphology from
comparisons of our models with Galactic radio observations, we really
only have \lvplot s!  Fig.~\ref{lvoftime} shows \lvplot s obtained
from the gas distributions in the first and last panel of
Fig.~\ref{rhooftime}, at $t=0.3\gyr$ and $t=3\gyr$, respectively, for
an assumed distance of the Sun to the galactic center of $8.0\kpc$ and
LSR rotation velocity $v_0=200\kms$. The bright ridge rising steeply
from the center in these diagrams is caused by the dense disk of gas
on $x_2$--orbits, visible in the very center of the flow in
Fig.~\ref{rhooftime}. The more irregularly--shaped ridges are the
traces of spiral arms in the \lvplot. Also well visible in
Fig.~\ref{lvoftime} are the terminal velocity curves.

Looking at Figs.~\ref{rhooftime} and \ref{lvoftime} shows that the
relation between morphological structures in the gas disk and
corresponding structures in the \lvplot\ is somewhat non-intuitive. In
order to gain a better understanding of this relation, we have
constructed a schematic representation of the arm structures of the
model in the top left panel of Fig.~\ref{rhooftime} in both coordinate
planes.  The upper panel in Fig.~\ref{lvexplain} shows schematically
the location of the gaseous spiral arms, the cusped orbit shocks (dust
lanes), and the $x_2$--disk in the ($x,y$)-plane. In this diagram, the
Sun is located at $x=-7.5\kpc$, $y=-2.7\kpc$, i.e. at $R_0=8\,\KPC$
and $\PHIBAR=20^\circ$.  The lower panel shows the corresponding
features in the \lvplot\ as observed from this LSR position, with
the same line styles to facilitate cross identification.

We see that whenever a spiral arm crosses a line--of--sight from the
Sun twice, it appears as a part of a loop in the \lvplot. This is the
case, e.g., for the two outer spiral arms seen nearly end--on (thin
full lines in Fig.~\ref{lvexplain}), and for the innermost pair of
arms driven by the bar (thick and thin dashed lines).  The equivalent
to the $3\kpc$ arm (see below) and the corresponding counterarm on the
far side of the galaxy are parts of a second pair of arms driven by
the bar; these cross the relevant lines--of--sight to the Sun only
once (thick full lines and thick dotted lines in
Fig.~\ref{lvexplain}), respectively). The same is true for the outer
pair of spiral arms seen nearly broad--on as viewed from the Sun
(dash--dotted and small dotted lines).

It is clear from inspection of Figs.~\ref{lvoftime} and
\ref{lvexplain} and a comparison with the corresponding observational
data (see the figures reproduced in Section 2, and the diagrams in the
papers cited there) that already the initial COBE--constrained model
gas flow of Fig.~\ref{rhooftime} resembles the Milky Way gas
distribution in several respects:

(i) The number of arm features in the longitude range
$[-60\deg,60\deg]$ and their spacing in longitude is approximately
correct (compare Table~\ref{armstable}).

(ii) The model contains an arm which passes through the $l=0$--axis at
negative velocity ($\sim -30 \kms$) and merges into the southern
terminal velocity curve at negative $l$.  Qualitatively, this is
similar to the well--known $3\kpc$--arm, although this crosses the
$l=0$--axis at $\sim -50 \kms$ and extends to larger longitudes.

(iii) The positions and velocities of gas particles in the $x_2$--disk
are similar to those observed for the giant Galactic Center molecular
clouds on the $l>0$ side in the CS line (Bally \etal 1987, 1988;
Binney \etal 1991).

(iv) The terminal velocity curve slopes upwards towards large
velocities near $l=0$, although not as much as would be expected for
gas on and just outside the cusped orbit in the COBE potential (see
\S~\ref{sec-termcurve}), and not as much as seen in the HI and CO
data. In the following subsections we compare the COBE models more
quantitatively with the observational data, and attempt to constrain
their main parameters.

\subsection{The terminal velocity curve}
\label{sec-termcurve}

\begin{figure*}
\ifpsfiles\psfig{figure=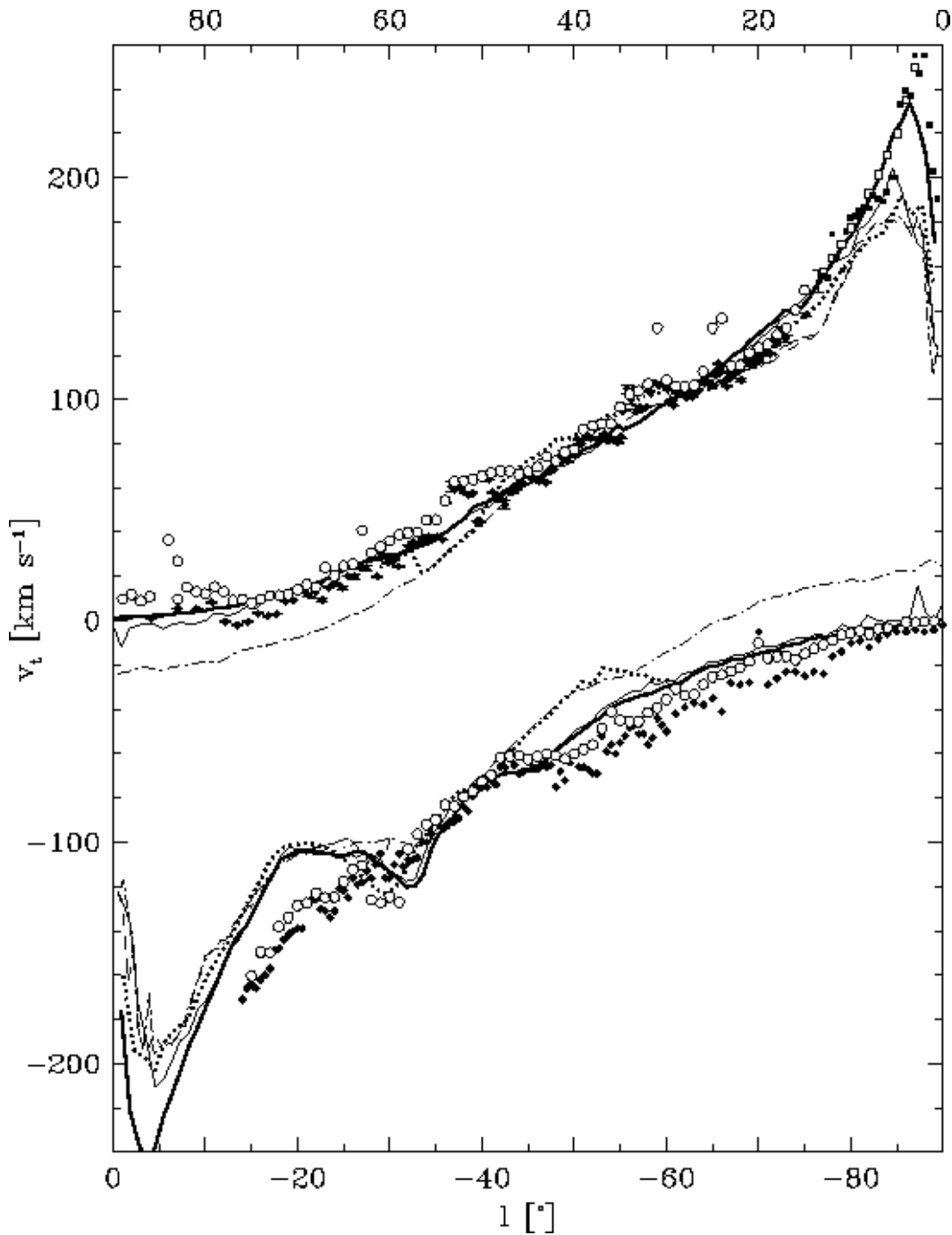,width=14cm}
\fi
\caption[]{%
Northern and southern Galactic terminal velocity curves compared with
model predictions. Observational data from sources as follows.  Filled
squares ($l=0-10\deg$): HI data from Fig.~1 of Burton \& Liszt
(1993).  Empty squares ($l=0-20\deg$): unpublished 140ft single dish
HI data, kindly provided by Dr.~B.~Burton.  Open circles: HI terminal
velocities from Fich \etal (1989), based on data from Westerhout
(1957). Diamonds, with tiny error bars: northern \twco\ terminal
velocities from Clemens (1985). Without error bars: southern \twco\
data from Alvarez \etal (1990). The data are relative to the LSR,
mostly as corrected by the respective authors.  The Clemens (1985)
data have been corrected for internal dispersion, and the velocities
in Burton \& Liszt (1993) are relative to the Sun and have been
corrected for LSR motion.

\hskip 10pt
The model terminal velocity curves are from: a typical low resolution
model with no halo and declining rotation curve (dash--dotted); the
same model at higher resolution but with less radial extent (dotted);
a low resolution model with flat rotation curve (thin solid); and our
best model with high resolution (see text) and flat rotation curve
(thick solid line). All models assume a bar angle $\phibar=20\deg$
and have corotation at $R_c\simeq 3.4 \kpc$.
}
\label{fig-term0.3}
\end{figure*}

\begin{figure}
\ifpsfiles
\psfig{figure=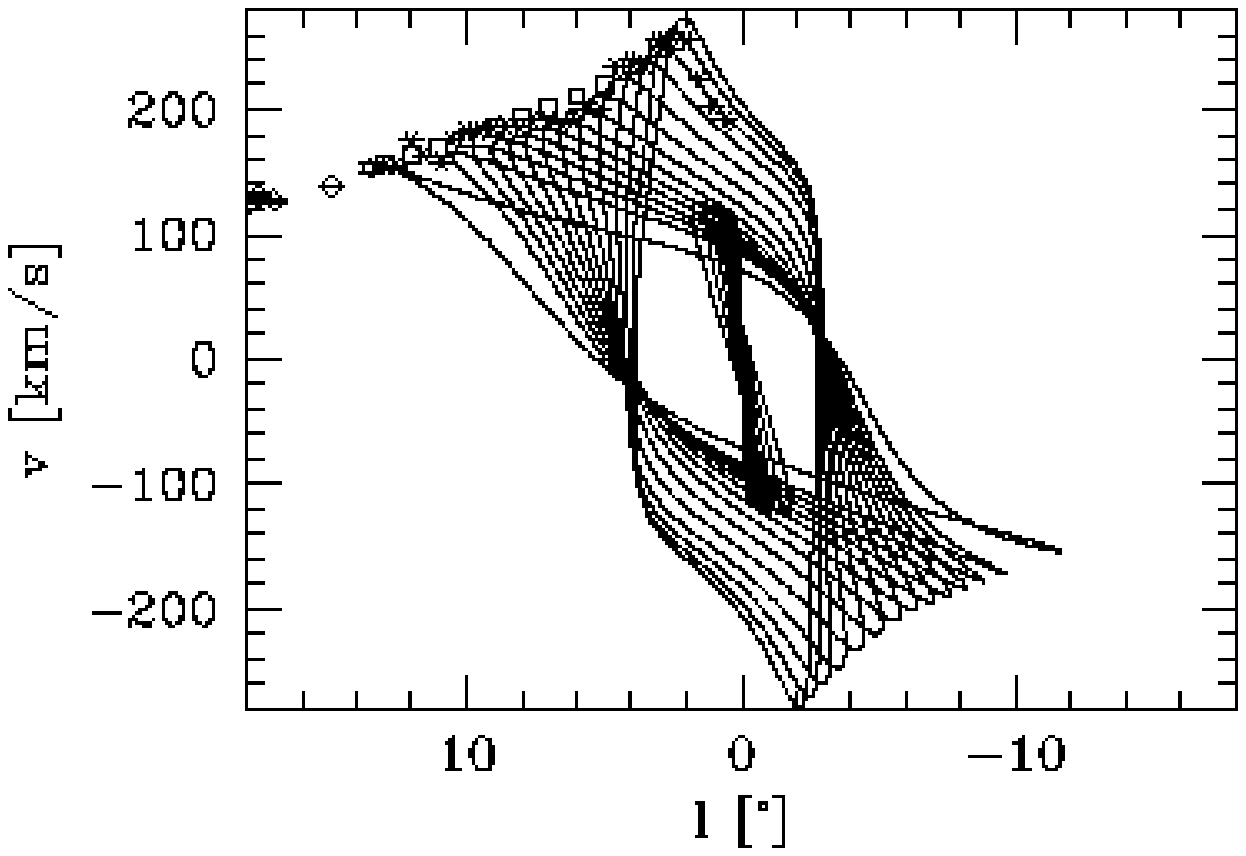,width=8.2cm}

\hspace{-0.28cm}\psfig{figure=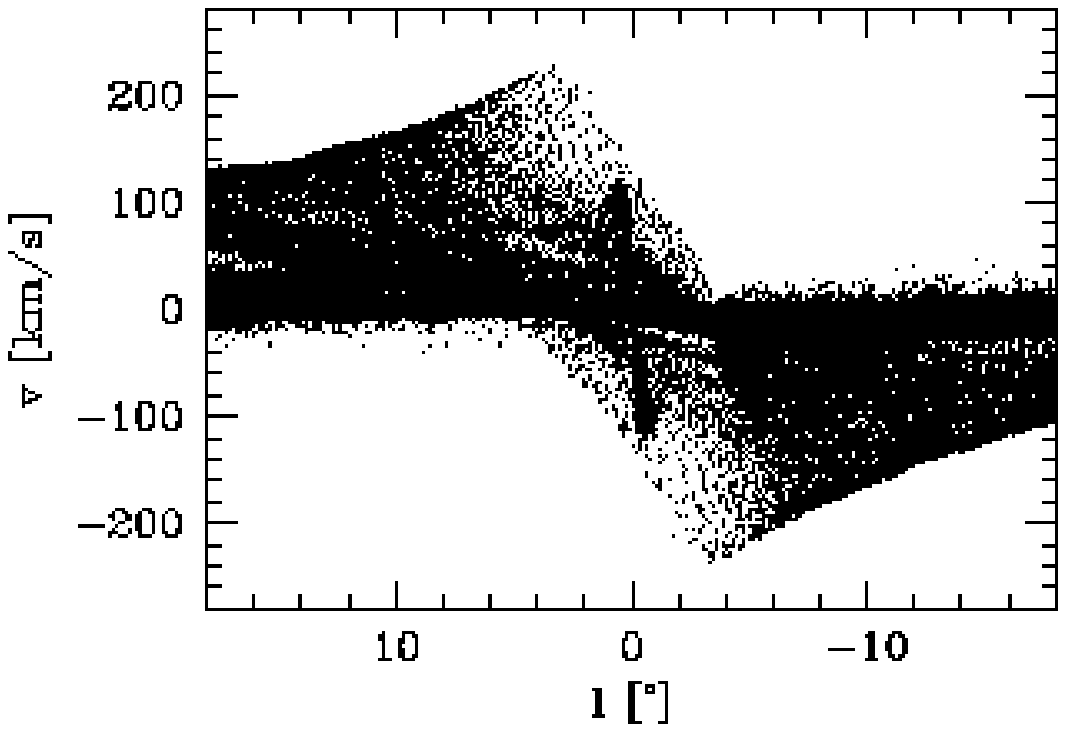,width=8.5cm}
\else\vskip6cm\fi
\caption[]{%
Top: The upper envelope of the $x_1$-orbits in the ($l,v$)-plane
traces out the observed inner terminal velocity curve. The figure
shows $x_1$-orbits computed in the same gravitational potential as
that used for the models shown in Fig.~\ref{fig-term0.3}; each
$x_1$-orbit traces a parallelogram--like curve in this plot. The
innermost $x_1$-orbit shown is the cusped orbit which peaks at
$l=2^\circ$. Inside this orbit, the gas switches to $x_2$-orbits which
reach velocities of $120\,\kms$ in this potential. The observed
northern terminal velocities are shown for comparison (same sources as
in Fig.~\ref{fig-term0.3}).

\hskip 10pt
Bottom: Particle distribution in the same part of the ($l,v$)-plane,
for the high--resolution model shown by the thick solid line in
Fig.~\ref{fig-term0.3}. The cusped orbit and the innermost few
$x_1$-orbits are not occupied even at this resolution. The
parallelogram structure in this plot therefore reaches inwards and
upwards to only $l\simeq3\deg$ and $v\simeq220\kms$, compared to
$l\simeq2\deg$ and $v\simeq270\kms$ for the orbits, and $l\simeq2\deg$
and $v\simeq260\kms$ for the observed terminal velocities.
}
\label{fig-orbitslv}
\end{figure}

We will now determine the mass normalisation of the models by fitting
their predicted terminal velocity curves to observations.  Model
terminal curves are computed by searching for the maximal radial
velocity along each line of sight, as seen from the position of the
Sun $8\kpc$ from the center, and then substracting the projected
component of the LSR motion.  For the latter we have assumed that the
LSR motion is along a circular orbit with $v_{\rm LSR}=200\kms$ and no
radial velocity component. By an eye-ball fit of these model curves to
the observed terminal velocity curve we then obtain the proper scaling
constant $\xi$ (see \S~\ref{sec-scaling-def}) by which all
velocities in the model have to be multiplied to obtain their Galactic
values. Both the mass and the potential then scale with $\xi^2$.

Figure \ref{fig-term0.3} shows several model terminal velocity curves
obtained in this way, and compares them with the northern and southern
Galactic terminal velocities as determined from HI and \twco\
observations from a number of sources. Although the error bars for all
measured terminal velocities are small, the data show some scatter due
to differences in angular resolution and sensitivity.  The model
terminal velocity curves Fig.~\ref{fig-term0.3} are for models of
different spatial resolution and simulated radial extent, to
illustrate the effects of these parameters. The dash-dotted line is
for a model extending to galactocentric radius $9\kpc$; this
demonstrates that for $v_{\rm LSR}\ge200\kms$ the observed terminal
curve beyond about $\pm (40-50)^\circ$ requires a dark halo component
in the Milky Way.

After including a halo component in the model (by modifying the
monopole component of the potential such that the rotation velocity
becomes constant outside $4.5\kpc$; see Fig.~\ref{rotationcurve}),
both the northern and southern terminal curves are much better
reproduced (thick solid line in Fig.~\ref{fig-term0.3}). The rotation
curve of this halo model is shown in Fig.~\ref{rotationcurve}; the
simulated radial range is $12\kpc$. The contribution of the dark halo
inside the solar radius is fairly small ($\sim 23\%$ in the radial
force at $8\kpc$), somewhat less even than in Kent's (1992) maximum
disk model.

In this model, there remain two main regions of discrepancy with the
observed terminal velocity curve. First, the model terminal velocities
are too low at and just outside the peak at $l\simeq 2\deg$. This is
strongly influenced by and probably due to resolution effects, as
discussed below. Second, there is a larger mismatch around
$-20^\circ$. This is likely caused by our mass model not being correct
in the vicinity of the NIR lumps $\sim 3\kpc$ down the minor axis of
the bar. Lines--of--sight at around $-20^\circ$ cross one of these
lumps as well as the end of the $3\kpc$ arm and the head of one of the
spiral arms outside corotation (see Fig.~\ref{lvexplain}). The
eight-fold symmetric deprojection of BGS is therefore likely to give
incorrect results in this region. Smaller systematic deviations in the
terminal velocities are visible around $l=30-50\deg$ and
$l=-(50-70)\deg$, although there, and everywhere else, the differences
between model and observations are now of the order of the scatter
between the various observational data and of the order expected from
perturbations in the disk. Given the uncertainties the overall
agreement is surprisingly good. This suggests that the basic
underlying assumption, that in the inner Galaxy the NIR light traces
the mass, is mostly correct.

Fitting the terminal velocity curve to both sides we obtain a scaling
constant of $\xi=1.12$.  This is slightly larger than the value
obtained in our first attempts to fit the models to the observations
($\xi=1.075$, Bissantz \etal 1997), in which we only considered the
northern rotation curve and ignored the data beyond $l=48^\circ$.  It
is worthwhile pointing out that the derived value of $\xi$ is only
weakly dependent on the assumed LSR tangential velocity: In order of
magnitude, a $10\%$ change in the LSR velocity leads to a $1\%$
difference in $\xi$. With an improved dust model and deprojection of
the outer disk we could therefore attempt to determine $\VSUN$ from
these models.

As is clearly visible in Fig.~\ref{fig-term0.3}, the peak in the
observed terminal curve at $2^\circ$ is not well reproduced by our
lower--resolution models. However, Fig.~\ref{fig-orbitslv} shows that
it {\sl is} nicely approximated by the envelope of the $x_1$--orbit
family when all orbital velocities are scaled by the same value of
$\xi$. At early times in the model evolution, when the gas flow is not
yet stationary, the peak is also reproduced in the hydrodynamic gas
model, but thereafter the region around the cusped orbit is
depopulated (see also Jenkins \& Binney 1994).  We attribute this to
the artifical viscosity in the SPH method, which smears out the
velocity gradient over two smoothing lengths, and to the method used
for setting up the gas simulation.

We can estimate the magnitude of the effect as follows.  Near the cusped
orbit, which sets the maximum velocity along the terminal velocity curve,
the particle smoothing length $h$ in the low resolution model is large,
about $\sim 100\pc$, because the gas density in this region is
small. The full $x_1$--orbit velocity on the terminal curve can only be
reached about two smoothing lengths away from the cusped orbit, where it is
no longer affected by the more slowly moving gas on $x_2$--orbits further
in.  The longitudinal angle corresponding to about $2h$ at the distance of
the galactic centre is about $1.4^\circ$. Thus, the peak in the gas
dynamical terminal curve should be found at $l > 3.4^\circ$ when the orbits
peak at $2^\circ$, showing how sensitive the peak location is to resolution.

To test this explanation, we have run a bi-symmetric model with
$100,000$ particles, resulting in about 2.2 times
as much spatial resolution as in the low--resolution models.
Just on the basis of this higher resolution, the terminal velocity
peak then moves from about $5^\circ$ to $4^\circ$ (at
$t=0.3\,\GYR$). We then further increased the resolution by the
following procedure. The gas inside the outermost $x_1$ orbit shown in
Fig.~\ref{resdiagram} was removed, and set up again on nested closed
$x_1$ and $x_2$ orbits, while keeping particles outside this region
unchanged. Evolving this modified gas distribution for a further
$0.3\,\GYR$, we obtained our final high resolution model. This is
shown by the thick solid line in Fig.~\ref{fig-term0.3}, which peaks
at about $3^\circ$ and $v_t=235\kms$. Compared to the original $20000$
particle model (thin solid line in Fig.~\ref{fig-term0.3}) the
mismatch at the peak has been reduced by about a factor of two in
scale and by two thirds in the peak velocity.

Although this analysis was inspired by a technical problem, there is
an observable implication of it as well. Since we may interpret the
hydrodynamical model in terms of gas clouds having a mean free path
length of order the smoothing length, we may restate the result in the
following way: a loss of resolution occurs when the cloud mean free
path is significant compared to the gradient in the true velocity
field. Applied to the inner Galaxy, our result then indicates that the
clouds near the cusped orbit peak in the terminal velocity curve must
have short mean free paths, i.e., be described well in a fluid
approximation.

Apart from resolution effects, the precise position of the peak in the
terminal velocity curve also depends critically on the location of the
ILR and hence on the mass model in the central few $100 \pc$. In this
region, the deprojected COBE model suffers from a lack of resolution
and our added nuclear component has uncertainties as well. We
therefore believe that with improved data and further work the
remaining discrepancies in this region will be fixed.

\subsection{Pattern Speed and Orientation of the Galactic Bar}

\label{sec-patternspeed}
\begin{figure*}
\ifpsfiles\hbox to \hsize{\vbox{
	    \centerline{\psfig{figure=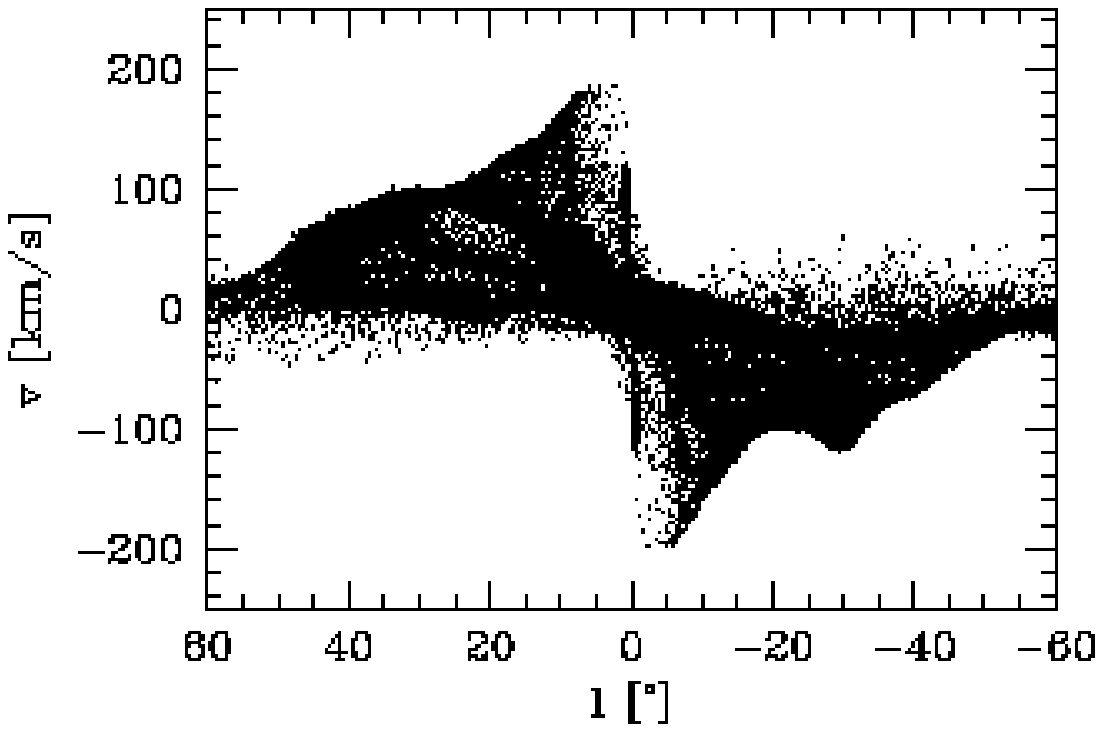,width=8.2cm}
	        \quad \psfig{figure=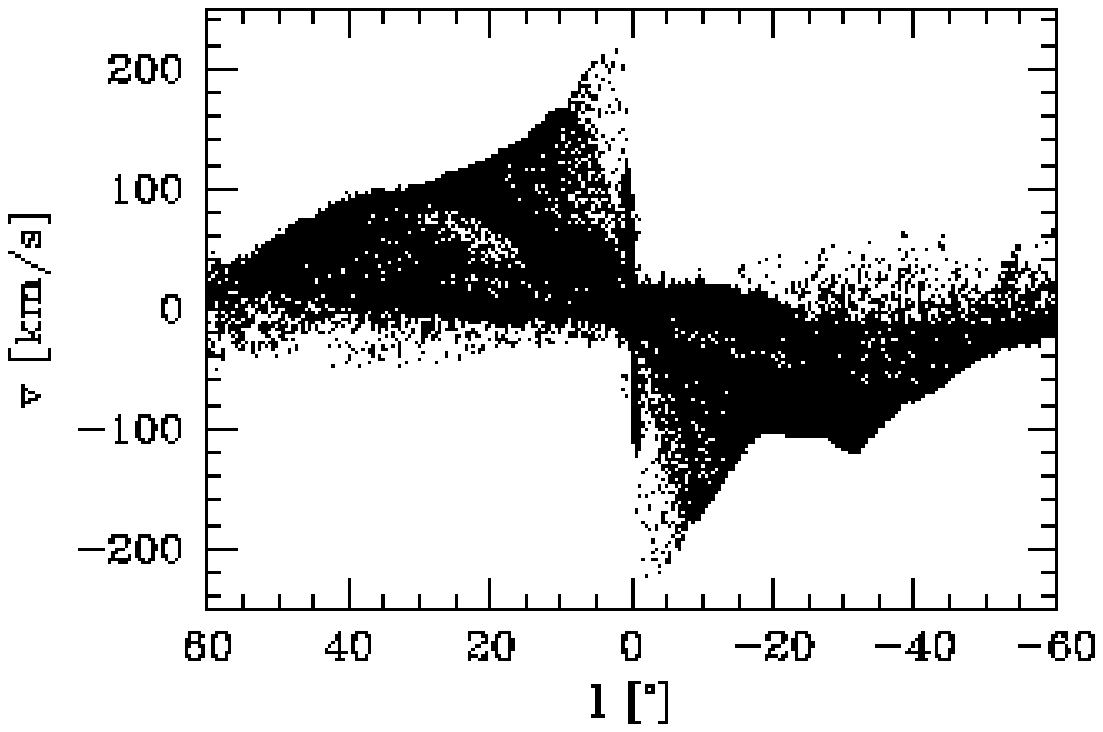,width=8.2cm}}
		\medskip
	    \centerline{\psfig{figure=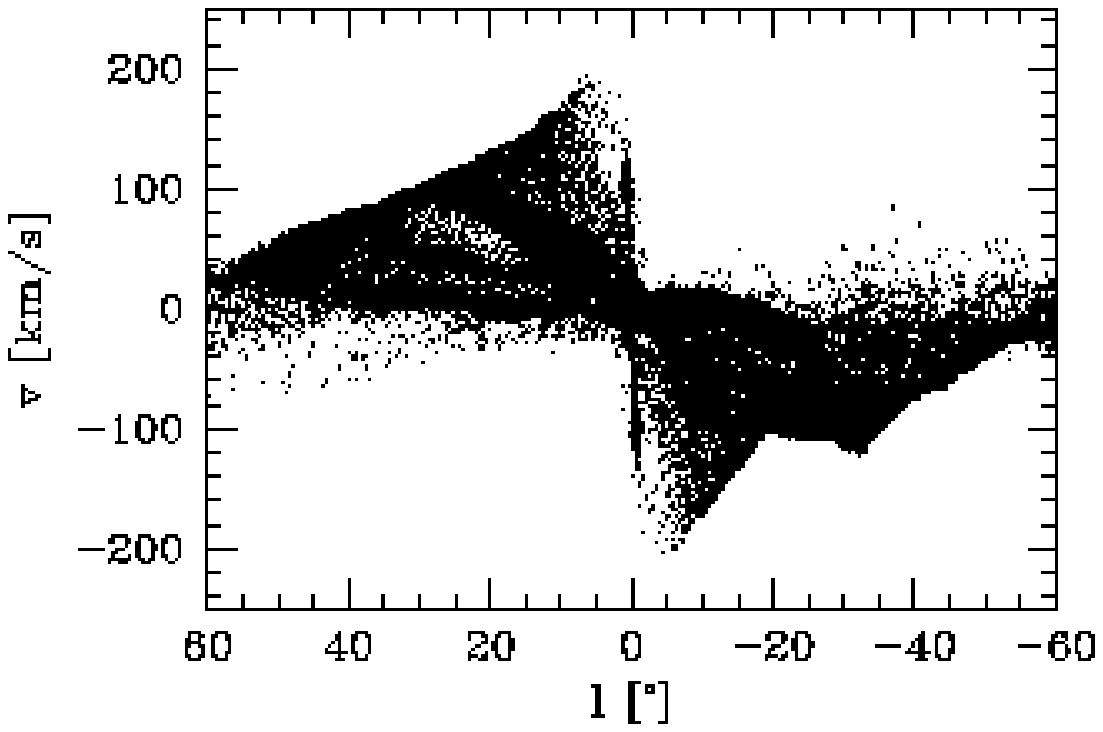,width=8.2cm}
	        \quad \psfig{figure=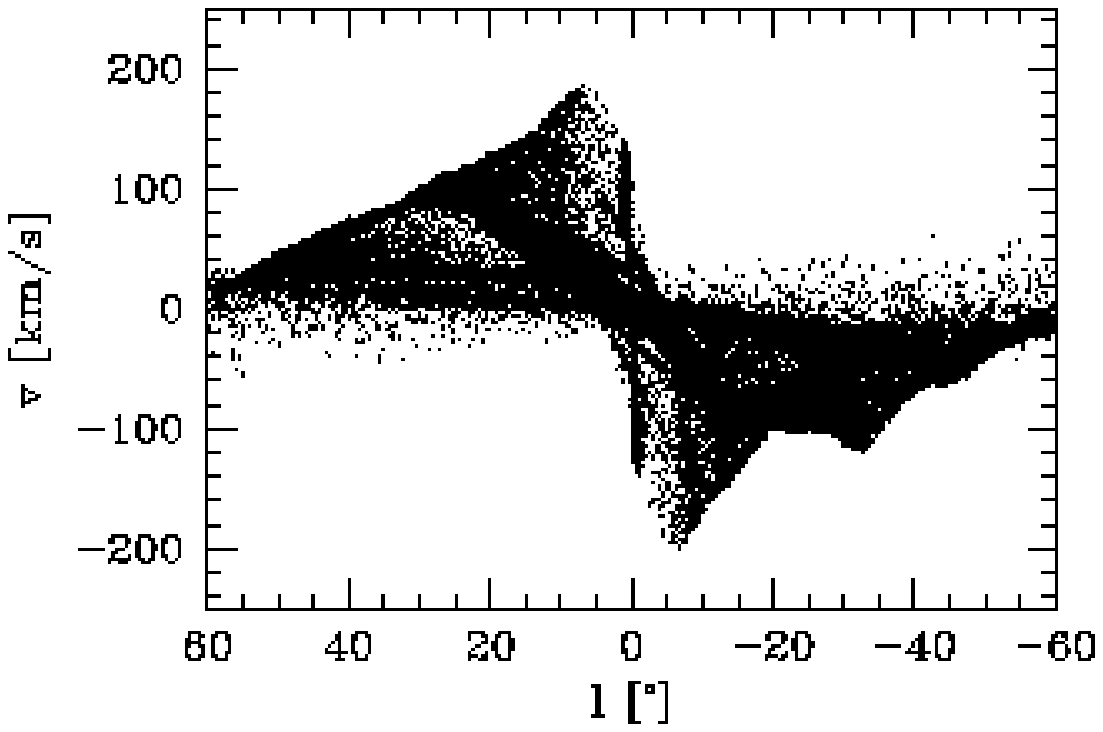,width=8.2cm}}
		\medskip
	    \centerline{\psfig{figure=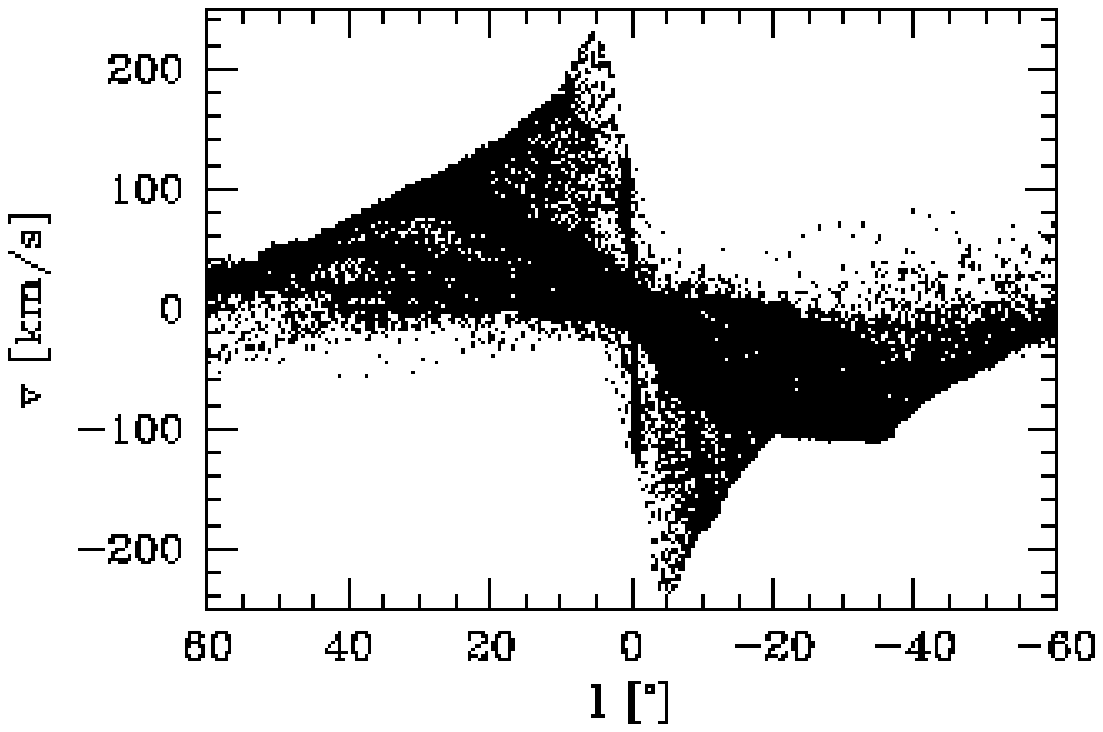,width=8.2cm}
	        \quad \psfig{figure=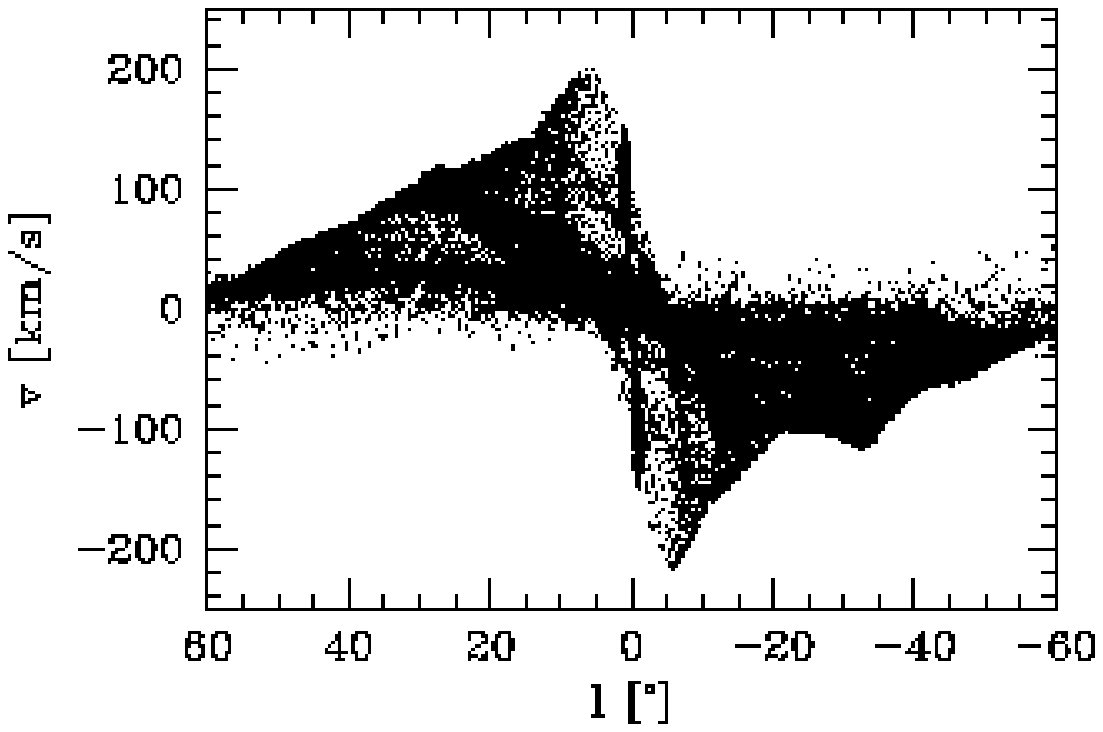,width=8.2cm}}
		} }
\else\vskip12cm\fi
\caption[]{%
Longitude--velocity \lvplot s for the gas flows in COBE bar potentials
with different pattern speeds and bar orientation angles. The column of
\lvplot s on the left shows the influence of the pattern speed
on the gas flow in the standard $\PHIBAR=20\deg$ COBE bar potential.
These frames  show models with corotation radii $R_c=3.1\kpc$ (top)
$R_c=3.4\kpc$ (middle) to $R_c=4.0 \kpc$ (bottom). The
right column shows gas flows in COBE bars deprojected for different
bar orientation angles, $\PHIBAR=15\deg$ (top), $\PHIBAR=25\deg$ (middle),
$\PHIBAR=30\deg$ (bottom), all for corotation at $R_c=3.4\kpc$.
}
\label{omegaphi}
\end{figure*}

There are two observations which constrain the value of the pattern speed
rather tightly. First, there is the 3-kpc-arm, a feature which exhibits
non-circular motions of at least $50\,\KMS$.  In our models, we find that
only the arms inside the bar's corotation radius are associated with strong
non-circular motions, so such an arm has to be driven by the bar.  From
observations and models of barred galaxies we also know that strong spiral
arms associated with both ends of a bar are common.  Therefore we conclude
that the 3-kpc-arm must lie inside the bar's corotation radius.

Second, a lower limit to the corotation radius is given by the inner
edge of the molecular ring. If the molecular ring were indeed a ring
such as induced by a resonance, it would be located near the outer
Lindblad resonance (e.g., Schwarz 1981). On the other hand, if it is
actually made of several spiral arms (Dame 1993, Vall\'ee 1995, this
paper), then the small observed non-circular velocities along these
spiral arms also show that these arms must be outside the bar's
corotation radius.  Solomon \etal (1985) find from the distribution of
hot, presumely shocked cloud cores, that the inner edge of the
molecular ring is at $R=4\,\KPC$.  The total surface density of
neutral gas also drops dramatically inside of $4\kpc$ (Dame
1993). From the IRT $2.4\,\MYM$ photometry of the galactic disk Kent
\etal (1991) concluded that there is a ring, or spiral arm, at about
$R=3.7\,\KPC$. Therefore we conclude that the bar's corotation radius
is inside $R=4\,\kpc$.

An independent argument for corotation falling somewhere between
$3\kpc$ and $4\kpc$ comes from the fact that the deprojected COBE bar
appears to end somewhere between $3\kpc$ and $3.5\kpc$ (BGS).  From
both N--body simulations and direct and indirect observational
evidence the corotation radius is usually found at between 1.0 and 1.2
times the bar length (Sellwood \& Wilkinson 1993; Merrifield \&
Kuijken 1995; Athanassoula 1992b).  In our models, a coration radius
between $3\kpc$ and $4\,\KPC$ corresponds to a pattern speed of $\sim
50-60\,\KMSKPC$.

We have run gas dynamical simulations with corotation at 4.0,
3.4, and $3.1\,\KPC$, to determine from observations which of these
values is most nearly appropriate. For the comparison with
observations it is important to notice that several other parameters
enter here, most importantly, the orientation angle of the bar, the
uncertain contribution of the dark halo to the outer rotation curve
and hence terminal velocities, and the LSR velocity.

We first fix the bar orientation angle at $\phibar=20\deg$, but will
vary this parameter later. The chosen value of $\phibar$ is in the
range allowed by the NIR photometry (BGS), it is favoured by the clump
giant star distribution as analyzed by Stanek \etal (1997) and by the
gas kinematical analysis of the molecular parallelogram by Binney
\etal (1991), and it meets the preference for an end--on bar in the
interpretation of the microlensing experiments.

In the last section we found that the Galactic terminal velocity curve
for $\vert l\vert \leq 45\deg$ is well--reproduced by the gas flow in
the maximum NIR disk model with constant mass--to--light ratio.
Moreover, even this maximum disk model fails by a factor of $\gta 2$
in explaining the high microlensing optical depth towards the bulge
(Bissantz \etal 1997), making it very difficult to further reduce the
mass in the intervening disk and bulge.  We can therefore confidently
assume a maximum disk model in the following and, to separate the
determination of the bar and halo parameters, we restrict the
comparison with observations to longitudes $\vert l\vert \leq 45\deg$.

Finally, we set the LSR rotation velocity to $\VSUN=200\kms$, in the middle of
the observed range (\S2.7). A $10\%$ difference in this parameter is not
very important for the comparison with the inner Galaxy gas
velocities.

Thus we begin by considering a sequence of models with varying corotation
radius $R_c$ and the other parameters fixed as just described.  For three
models with $R_c= 4.0\kpc$, $3.4\kpc$, and $3.1\,\KPC$ we have plotted (l,v)
diagrams and have determined the scaling constant $\xi$ for each simulation
by fitting to both terminal curves. The final scaled \lvplots\ are shown in
the left column of Fig.~\ref{omegaphi}.  For the scaling constant we obtain
$\xi=1.13$, 1.12, and 1.09 for the 4.0, 3.4, and $3.1\kpc$ models. The
correctly scaled pattern speeds are then $59$, $57$, and $61\,\KMSKPC$. This
means that by changing the corotation radius, we effectively change the mass
of the model galaxy, while keeping the pattern speed almost constant at
about $60\,\KMSKPC$.

At small absolute longitudes, $\vert l\vert \leq 10\deg$, these
low--resolution models do not have enough particles to resolve the
true gas flow, and furthermore there are no published terminal
velocities in this region on the southern side.  Thus for now we
ignore data near the peak of the terminal velocity curve.  This leaves
a range of $\pm l = 10-45^\circ$ within which we compare these
no--halo model gas flows with the northern and southern terminal
velocity curves and with the various spiral arm features shown in
Figs.~\ref{obslvplot}--\ref{lvtracers}.

On the northern side ($l>0$), we try to match the model to the pronounced
spiral arm at about $+30^\circ$, which is best visible in the warm CO clouds
(Solomon \etal 1985) and in the distribution of HII-regions.
Moreover, the (l,v)-diagrams of CO and HII-regions shows that the
$+30^\circ$ arm is double (Fig.~\ref{lvtracers}).  In the model shown
schematically in Fig.~\ref{lvexplain}, there are actually three arms near
$l\simeq 30\deg$, two of which overlap, while the third, the northern
3-kpc-arm (thick dotted line), runs almost parallel to the first two. There
is also a wiggle in the terminal curve at about $+10^\circ$, which is
probably caused by a spiral arm similar to the northern, secondary inner
spiral arm in the model (thin dashed line in Fig.~\ref{lvexplain}).  The
southern terminal curve is more distorted by spiral arms than the northern
curve. A pronounced feature is the $-30^\circ$ arm in the molecular ring, as
well as the well--known 3-kpc-arm which continues on from a non--circular
velocity ridge beginning at $l\simeq 10\deg$ and $v=0$.

Fig.~\ref{omegaphi} (left column) shows that the gas flows in all three
cases are similar; nonetheless small differences in the spiral arm
locations help to show that the $3.4\,\KPC$ case is closest to the real
Galaxy.  The $30\deg$ spiral arm tangent is best reproduced in the
$3.4\kpc$-model (middle panel in left column of Fig.~\ref{omegaphi}). In the
top panel, it is not double as observed, and in the bottom panel the tangent
moves out to $\sim 40\deg$.  The $50\deg$ spiral arm tangent is reasonably
well reproduced in the top two panels, but is absent or very weak in the
bottom panel, but this arm may not be a reliable indicator in the absence of
a halo. In the south, we observe that large corotation radii move the arm at
$-30^\circ$ outwards. The $-30\deg$ spiral arm tangent is in about the
correct location in the top two panels, but too far out in the bottom panel.
The mismatch of the terminal curve at about $-20^\circ$ becomes larger with
decreasing corotation radius as well. On the other hand, the $-50\deg$
spiral arm tangent is not present for the largest pattern speed (top panel),
while it is adequately present in the lower two panels. The 3-kpc-arm
equivalent is present in all three models, but its detailed locus in the
\lvplot\ is not correct in any of the models: it has either too low
non--circular velocities at $l=0$ (particularly in the bottom panel), or it
does not extend to large enough negative longitudes (particularly in the top
panel). For corotation at $3.4\kpc$, the 3-kpc-arm ends in the model exactly
at $3\,\KPC$, as seen from the assumed solar position. There are particles
in the low intensity forbidden velocity region bounded by the line from
$(l,v)=(10,0)-(0,-50)$ in all three models, but the resolution does not
suffice to prefer one model over the other. Finally, an equivalent to the
$155\kms$ arm at small negative $l$ is present only for the lowest pattern
speed.

From an unweighted average of these comparisons to various observational
landmarks we conclude that the corotation radius of the Galactic bar is most
likely at about $3.4\,\KPC$. However, since none of the above models is
exactly right yet, this value could well change by $10\%$ when other effects
are taken into account, such as bar orientation (discussed next), dark halo
(\S\ref{sec-tangents}), or self--gravitating spiral arms
(\S\ref{sec-gravitating}).

With the corotation radius fixed at $3.4\,\KPC$, we can attempt to find an
optimal orientation angle for the bar. For this purpose, a series of
deprojected bar models have been made as in BGS, for bar orientation angles
$\phibar=10$, 15, 20, 25, and $30^\circ$. Smaller or larger angles are not
consistent with the asymmetry pattern of the observed NIR distribution or
result in unphysical bar shapes, see BGS and Bissantz \etal (1997).  With
the bar orientation the shape and radial extent of the deprojected bar
change.

Gas flow simulations with $R_c=3.4\,\KPC$ were made for each of these
cases, and \lvplots\ for $\phibar= 15$, 25, and $30^\circ$ are shown
in the right column of Fig.~\ref{omegaphi}. We observed that the
resulting changes in the model terminal velocity curves are caused in
about equal parts by the change in the viewing direction relative to
the bar, and by the intrinsic differences between the deprojected mass
distributions. The models' scaling factors $\xi$ are again determined
separately for each model by eyeball fitting the observed northern and
southern terminal curves. The variation in $\xi$ is very small,
however: we obtain $\xi=1.1$, 1.11, 1.12, 1.11, and 1.12 for orientation
angles from $10\deg$ to $30^\circ$.

We find only a weak preference for the $20^\circ$ model. The $-30^\circ$ arm
is more consistent with the $10\deg$ and $15^\circ$ case, whereas these
models show too large terminal velocities around $+40^\circ$. The
$+30^\circ$ arms seem to be in favour of $15-25^\circ$, whereas the
3-kpc-arm, although always too slow at $l=0$, seems to fit slightly better
with $20^\circ$. We conclude that the orientation is about $20^\circ$ with a
large uncertainty.

\subsection{Spiral arm tangents}

\label{sec-tangents}

\begin{figure*}
\ifpsfiles\hbox to \hsize{\vbox{
  \centerline{\psfig{figure=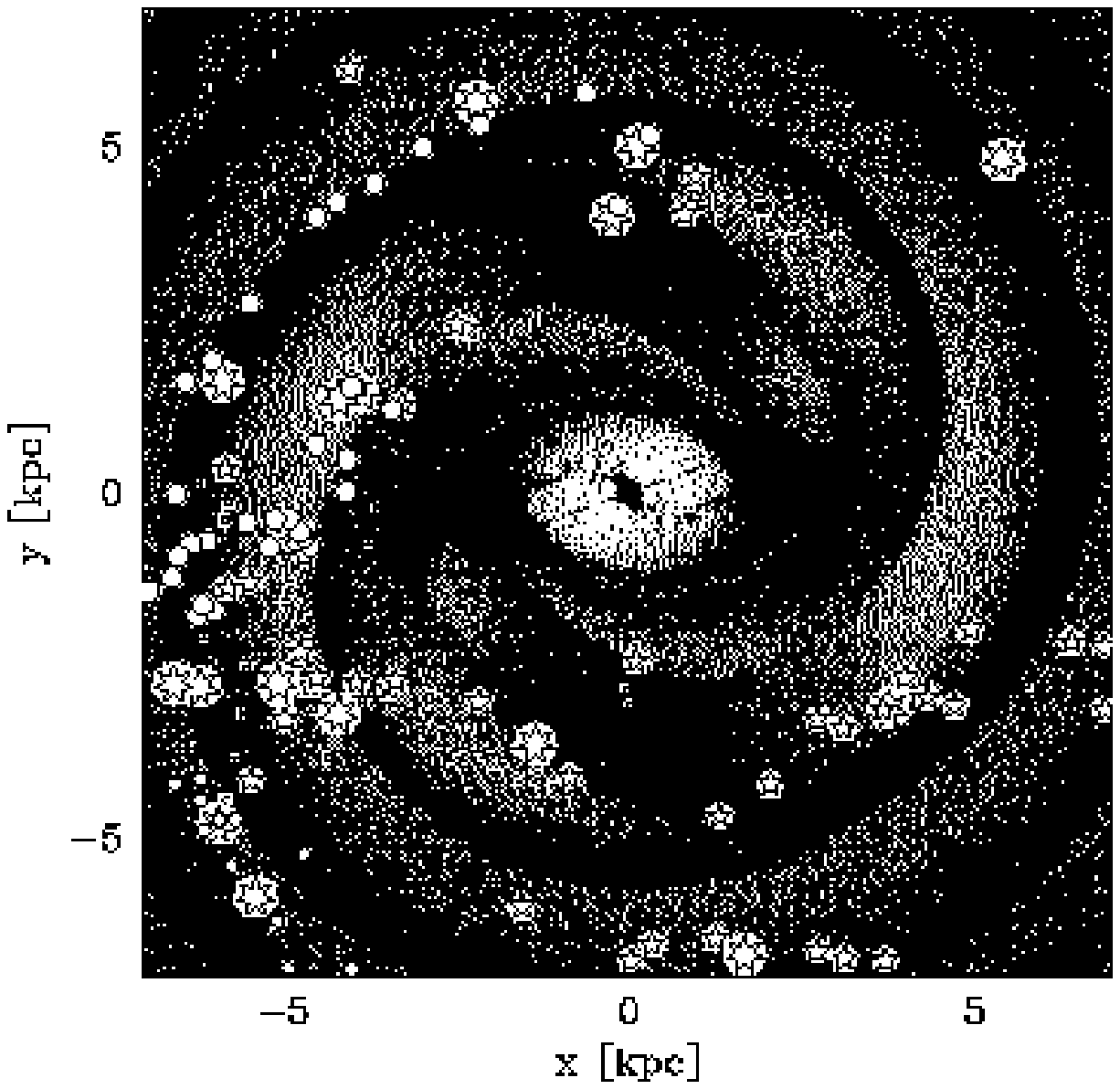,width=8.2cm}
        \quad \psfig{figure=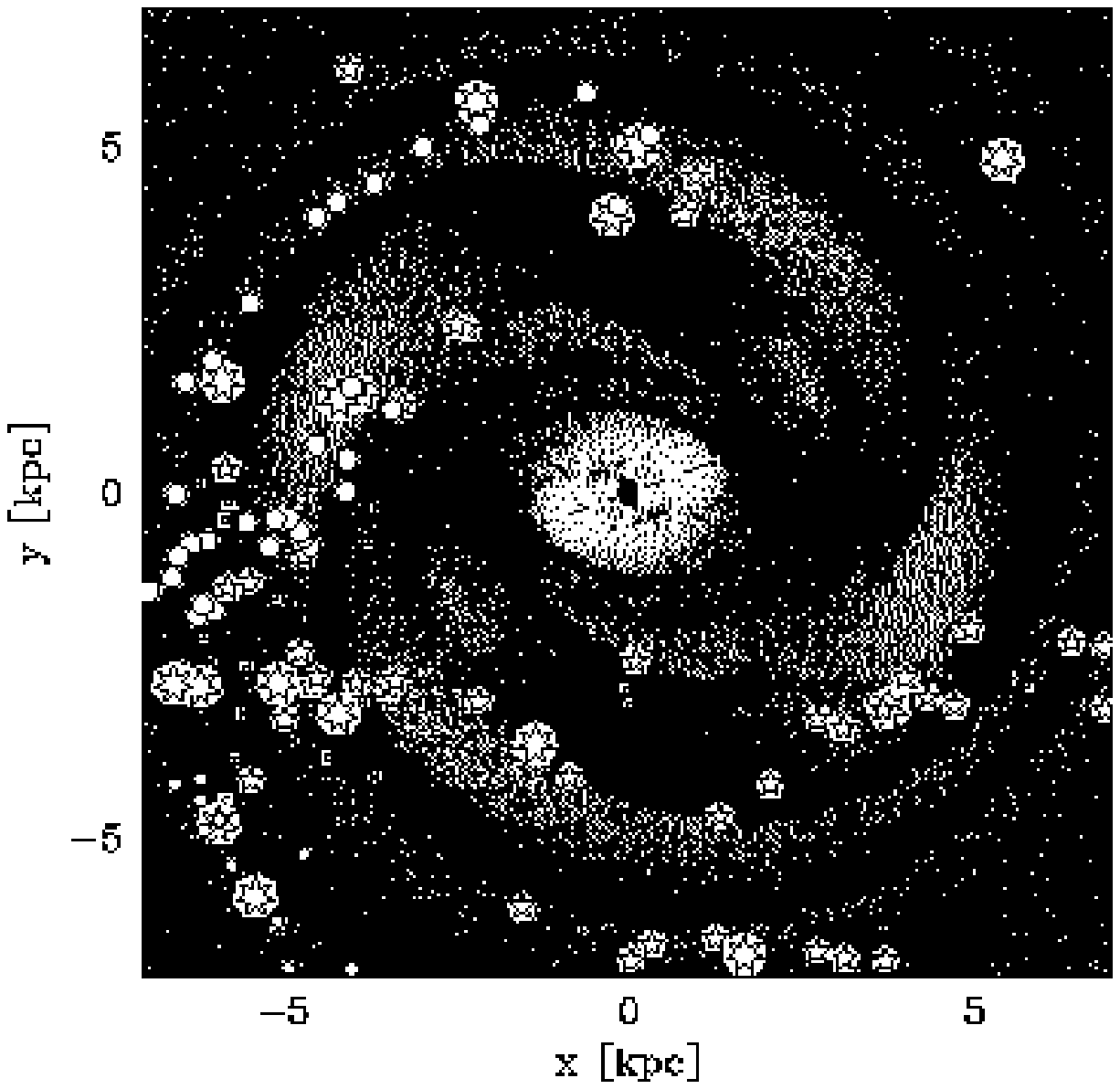,width=8.2cm}}
                } }

\hbox to \hsize{\vbox{
               \centerline{\psfig{figure=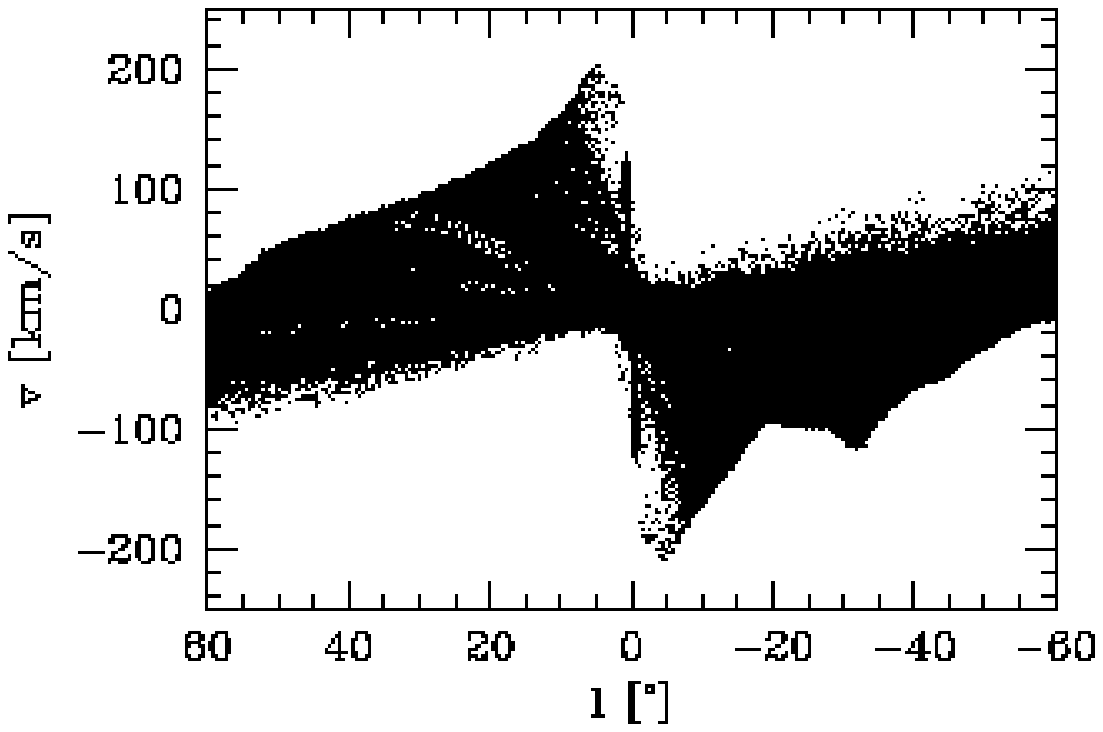,width=8.2cm}
                \quad \psfig{figure=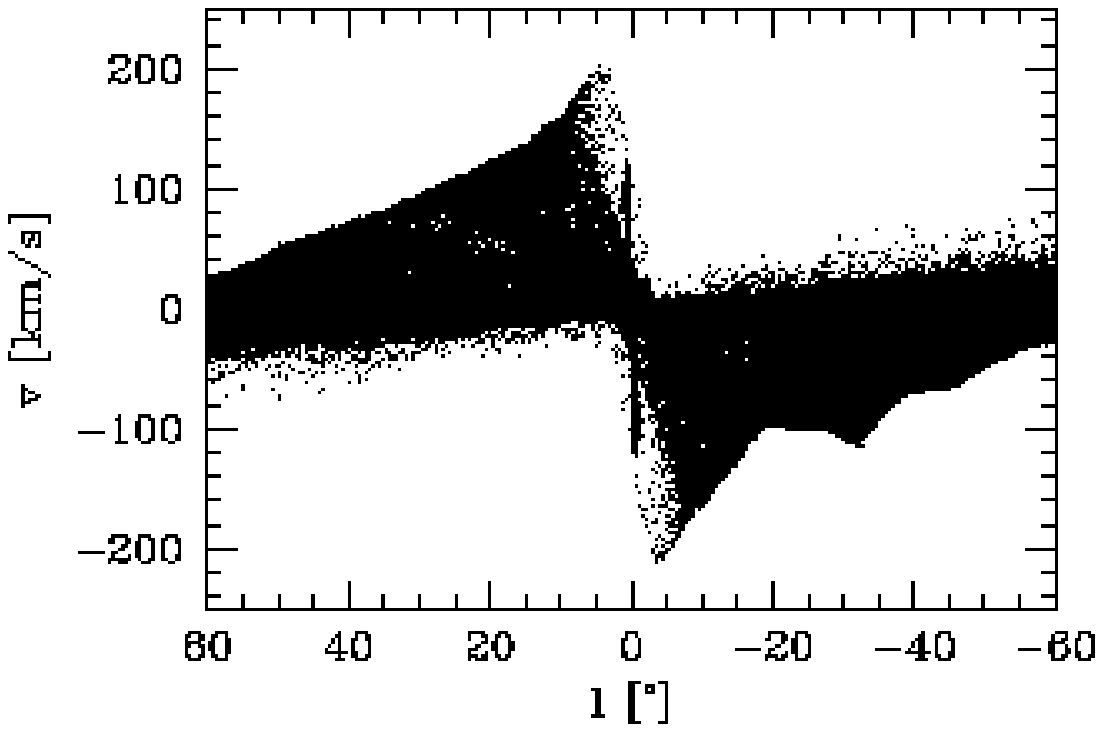,width=8.2cm}}
                } }
\else\vskip12cm\fi
\caption[]{%
Spiral arm tangents in two models with $\PHIBAR=20\deg$ compared to
observations.  Left: standard model without dark halo; right:
including dark halo.  Both models are computed with 100,000 particles
and the gas disk is truncated at $10\kpc$.

The observed directions of spiral arm tangents are shown by the
straight lines, starting at the position of the Sun at $x=-7.5\kpc$,
$y=-2.7\kpc$ in this plot. See Table~\ref{armstable} which also lists
the spiral arm tangents for the models.  For illustration, stars
denote positions of major HII-Regions from Georgelin \& Georgelin
(1976) and Georgelin \etal (1996), and circles the positions of large
molecular clouds from Dame \etal (1986) and Grabelsky \etal (1988).
The distances of these tracers have been rescaled to $R_0=8\kpc$ but
are {\sl not} corrected for non-circular motions.  Such a correction
would tend to move HII-Regions into the model spiral arms, because of
velocity crowding.
}
\label{fig-tang}
\end{figure*}

In the previous \S\ref{sec-patternspeed} we have already used the observed
spiral arm tangents at $l=\pm30\deg$ to constrain the pattern speed and
orientation angle of the bar.  Here we reconsider the location of the spiral
arms in the models that compared best with the Galaxy in
\S\ref{sec-patternspeed} (i.e., those with $R_c=3.4\kpc$ and
$\phibar=20\deg$), but with (i) a possible dark halo included, (ii) emphasis
also on the nearby terminal curve and the spiral arms with tangents at
$l=\pm50\deg$, and (iii) higher resolution.

For comparing the model spiral arms with observations we again use the
tracers in Fig.~\ref{lvtracers} (HII regions and molecular clouds) and the
characteristic features in \lvplot s like Fig.~\ref{obslvplot}.
Unfortunately, because the structures in the observed \lvplot s are much
less sharp than in the model \lvplot s, it is very difficult to measure
reliably any features beyond those already discussed, i.e., the spiral arm
tangent point positions, the 3--kpc arm and the molecular ring. Already the
run of the arms out of the molecular ring to their tangent points cannot be
identified unambiguously from Fig.~\ref{obslvplot}.

Spiral arm tangent directions in the models are easily determined.
When the arm is broadened in the tangential direction, we
place the tangent at the outer edge, where the velocity jump is. In this way
we can achieve a fair accuracy of a few degrees. Only one tangential
direction, the Scutum arm at $l=30^\circ$, cannot easily be determined in
this way, because its tangent goes through the corotation region through
which no arm continues inwards in the model.  Nevertheless, we can measure
an approximate value for this tangent as well. Table~\ref{armstable} gives a
comparison of model and observed spiral arm tangent directions, and the
various data that have been used for the observed directions.

The bar--driven spiral arms in the models change their relative strength,
opening angle, and hence location, depending on pattern speed, bar
orientation, halo mass contribution and other parameters.  In
\S\ref{sec-patternspeed} we have discussed the constraints
on the the pattern speed and bar orientation angle.
Fig.~\ref{fig-tang} shows that adding a dark halo mass contribution so
as to make the outer rotation curve flat (at $208\kms$ after scaling)
has the following two effects: (i) the outer pair of spiral arms at
$l=\pm50\deg$ now form a more regular pattern with the inner pair at
$l=\pm30\deg$, and all four arms now have a similar density contrast
with respect to the interarm gas; (ii) all four tangent directions now
fit the observations reasonably well; see also Table 1.
Fig.~\ref{fig-tang} also shows the positions of large HII-regions
and molecular clouds superposed on the gas arms.  Note that their
distances were determined from a circular orbit model and so may be
slightly in error.  Nonetheless it is reassuring that they fall
approximately on the model gas arms, consistent with the model's match
to the spiral arm tangents.

\subsection{Gravitating spiral arms}

\label{sec-gravitating}

\begin{figure*}
\ifpsfiles\hbox to \hsize{\vbox{
               
\centerline{\psfig{figure=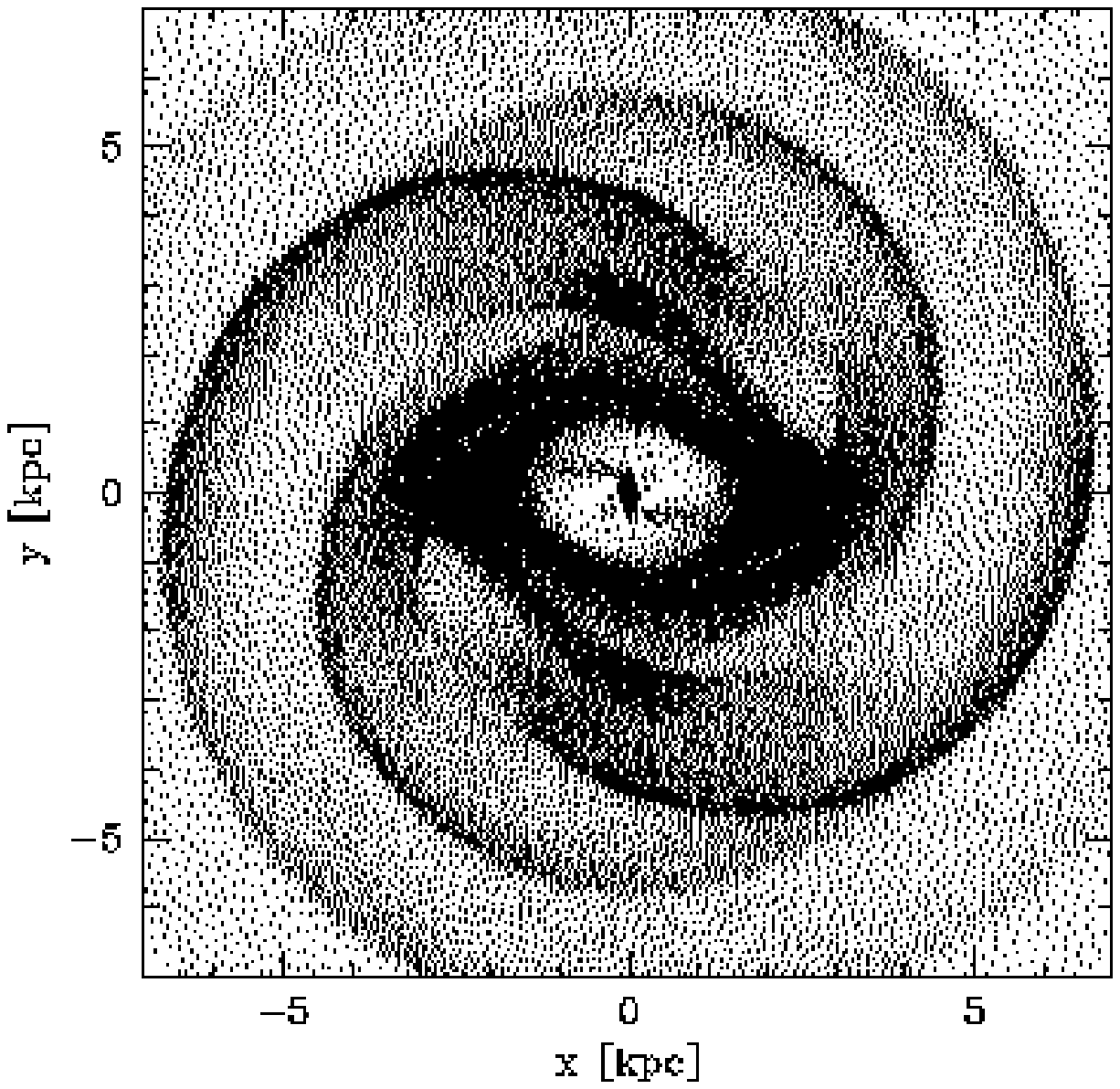,width=8.2cm}
                \quad \psfig{figure=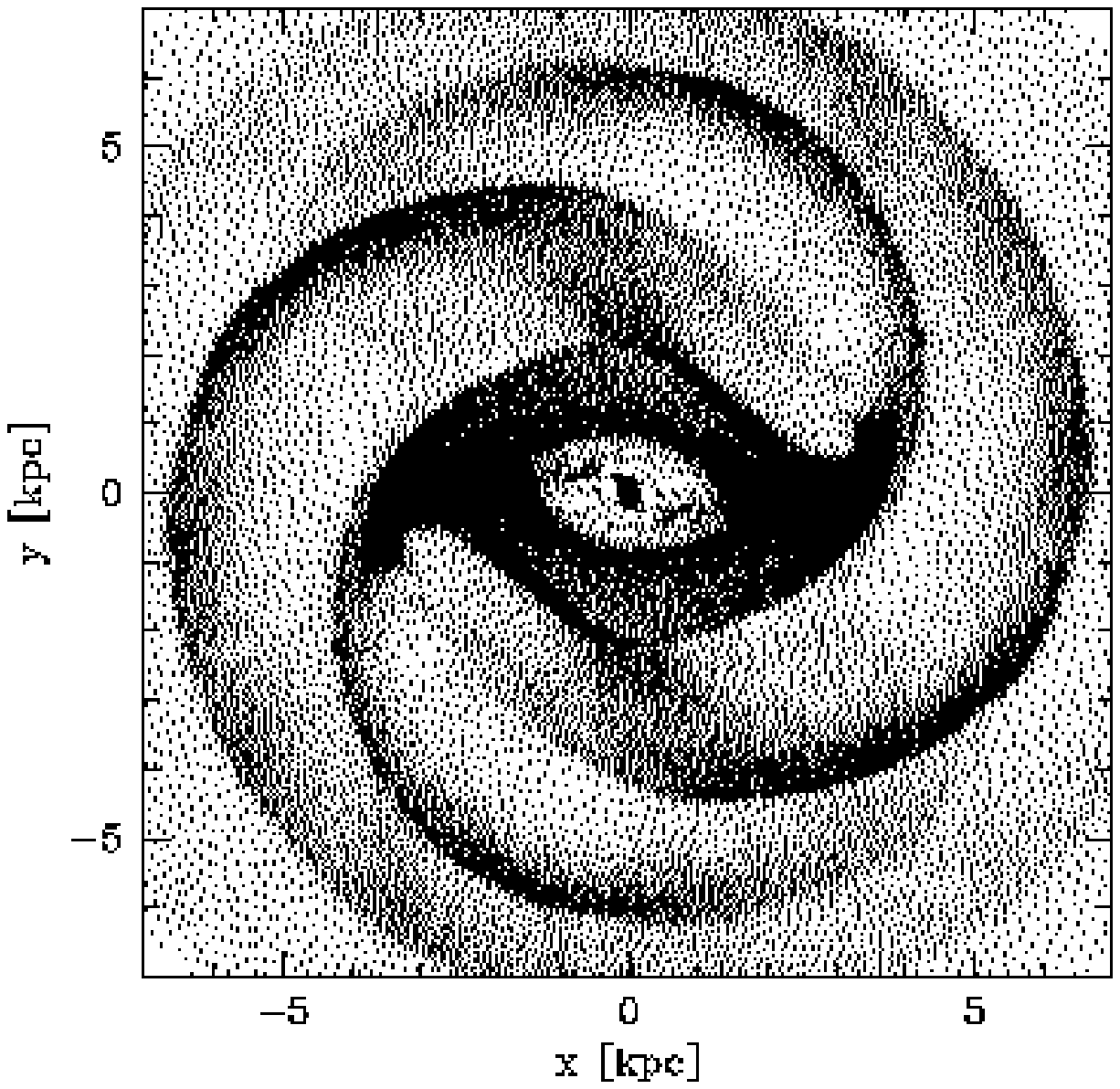,width=8.2cm}}
                } }
\hbox to \hsize{\vbox{
               
\centerline{\psfig{figure=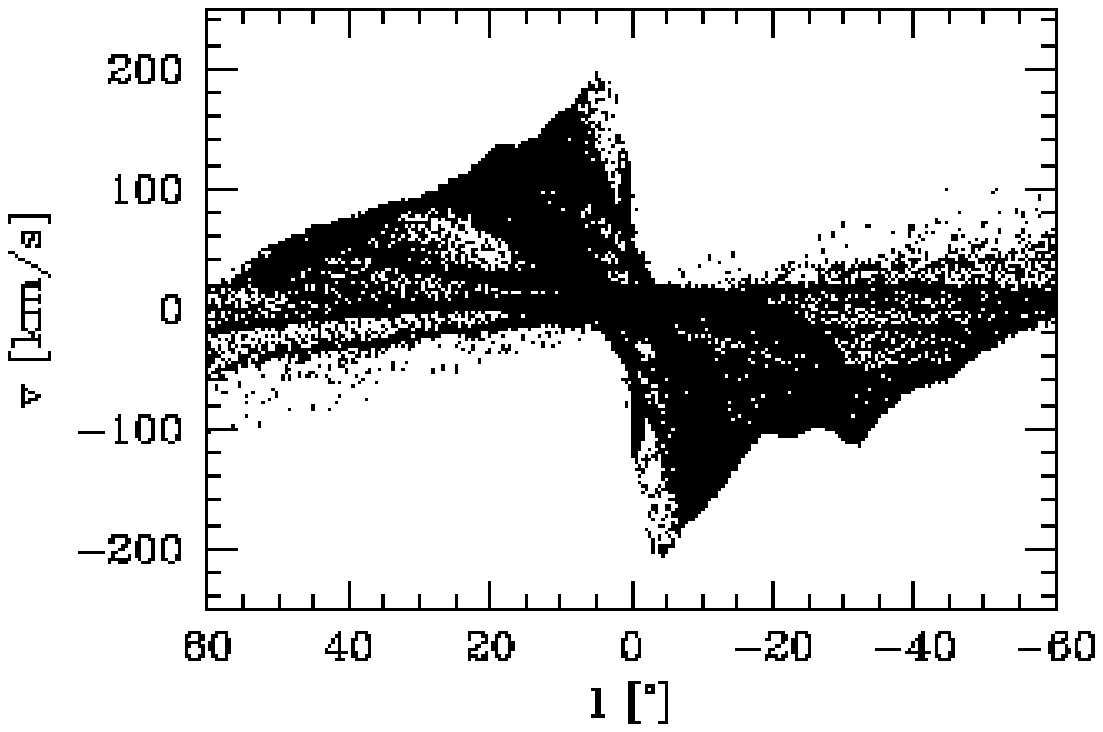,width=8.2cm}
                \quad \psfig{figure=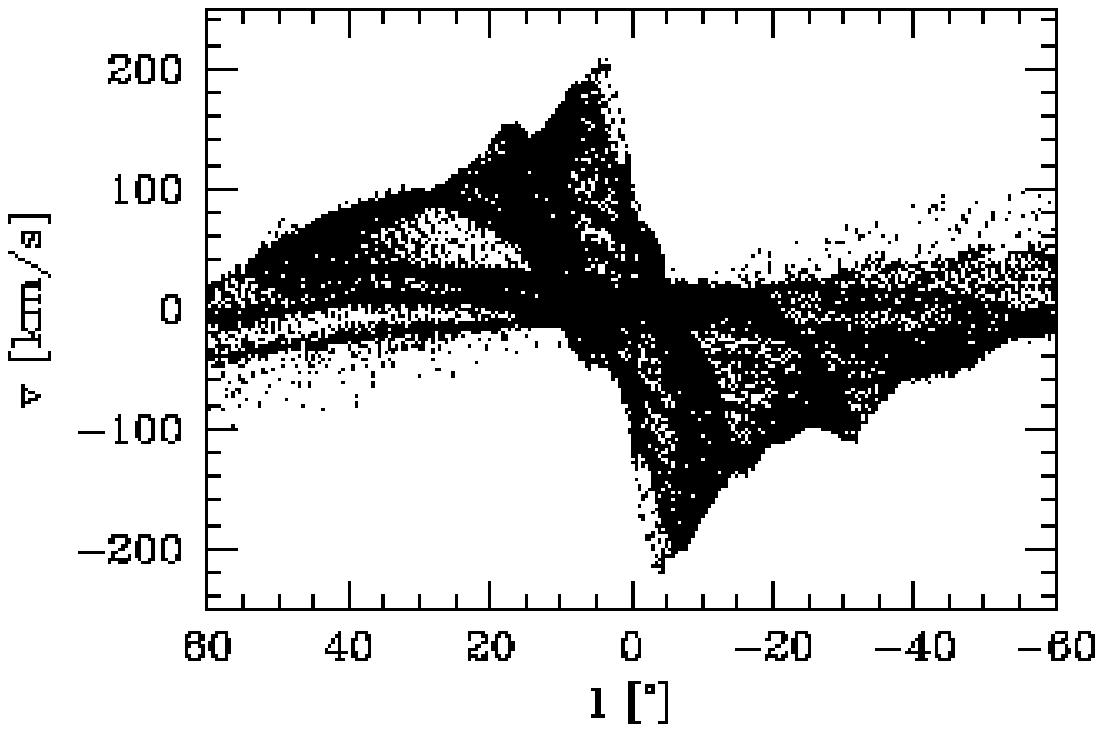,width=8.2cm}}
                } }
\else\vskip12cm\fi
\caption[]{%
Model with additional spiral arm potential. Left: case with
realistic parameters; right: twice as high surface density, and half as
large gravitational smoothing lengths.
}
\label{fig-grav}
\end{figure*}

In this section we estimate the effect, if any, of the gravitational
potential of the stellar spiral arms that are likely associated with
the spiral arms seen in the gas. Remember that in our model, the
spiral arms outside the bar's corotation radius are driven by the
clumps of NIR light and mass $\sim 3\kpc$ down the minor axis, which
rotate with the bar (see \S\ref{sec-morph}). We interpret these clumps
as the signature of real spiral arms in the deprojected NIR light. In
the gas model two of the spiral arm heads are about at the correct
positions where these clumps are observed; this supports the view that
the gravitational potential of the clumps is a first estimate of the
true spiral arm potential.

However, the models discussed sofar did not take into account the
gravitational perturbations outside the clump regions that could be
associated with the spiral arms.  In particular, we are interested in
knowing whether the morphology of the spiral arms would be changed
when these arms carry a reasonable fraction of the mass throughout the
disk.  To test this, we have experimented with the following scheme
that makes use of the particular strengths of the SPH method: First,
we assign a fraction of the mass of the NIR disk to the spiral
arms. Most of this mass will be associated with the spiral arm
perturbation in the stellar density. Then we assign this part of the
mass to the gas particles, assuming that a large fraction of the gas
will later be concentrated in the arms and that the shock fronts seen
in the gas will trace the stellar spiral arm crests. Finally, to mimic
the fact that the stellar spiral arms are much broader in azimuth than
the shocks seen in the gas, we set the gravitational softening radius
of the gas particles to a value appropriate for the stellar arm widths
and different from the smoothing length used in the calculation of the
hydrodynamic forces.

According to Rix \& Zaritsky (1995), the spiral arms cover about one
third of the surface area in galaxies morphologically similar to the
MW. The arm--interarm contrast is somewhat less than average for these
galaxies, so that we estimate that about $20\%$ of the total NIR
luminosity is in the spiral arms (superimposed on the axisymmetric
background disk which thus contributes about $80\%$).

The azimuthally averaged surface density associated with the spiral
arms should thus also be $\sim 20\%$ of the mean background stellar
surface density. In the solar neighborhood, the stellar surface
density of our normalized models is $35\xi^2\msun\pc^{-2}$.  Thus we
first take an `arm' potential corresponding to a mean surface density
of $7\msun\pc^{-2}$ at the solar radius, and a factor
$\exp[-(r-8\kpc)/2.5\kpc]$ more further in. Here we have used the
radial scale-length of the NIR disk from \S3. This spiral arm mass is
given to the gas particles, and in order to compensate for the extra
mass, we substract the same amount as a constant fraction from the
COBE NIR mass model.

Because the stellar spiral arms are less sharp than gaseaus arms, we
estimate their gravitational force by smoothing the
gravitational forces of the gas particles over some length scale
$\varepsilon$.  Each particle contributes a smoothed potential \eq{
\phi_i(\vek r)=-{G m_{\ast i}\over \sqrt{(\vek r-\vek
r_i)^2+\varepsilon^2}} } to the gravitational field of the
`arms'. Here, $m_{\ast i}$ denotes the stellar mass associated with
particle $i$; $m_{\ast i}$ is obtained by dividing the total mass in
the stellar spiral arms by the effective number of gas particles. The
parameter $\varepsilon$ must depend on the average distance between
two arms in the model (about $3\kpc$ in the region of interest). We
thus mimic the broader spiral arm potential by smoothing the gas arm
potential over about $\varepsilon \sim 1\,\kpc$.

Notice that the hydrodynamical forces do not depend on the actual
value of the surface mass density, but only on the particle 
density gradient.  We can therefore save some memory space by taking
the SPH particle mass $m_i$ equal to the stellar $m_{\ast i}$.

The initial gas disk is set up in a similar way as for the models
discussed above, but for consistency the surface density is that of an
exponential disk with the same radial scale length as the stellar
disk.  The resulting additional radial pressure gradient is far to
weak to change the dynamics.

In our first attempt at such a model we found that the mass in the gas
particles accumulating on the $x_2$-disk due to inflow was so large
that the rotation curve in this region was changed
significantly. Moreover, the dust lane shock fronts acquired
significant mass. This is unrealistic since no strong stellar arms
form at the edge of the bar, and it has the effect of changing the gas
flow near the cusped orbit. To avoid both effects we have set the gas
particle masses to zero inside the bar region, which was approximated
by an ellipse with major axis $3\kpc$ and minor axis
$1.8\kpc$. Correspondingly, the COBE mass model was then not changed
in this region.

The gravitating arm model with the specified parameters that finally
results is shown in the left panels of Fig.~\ref{fig-grav}.  First, we
notice that the distribution of gas has not changed much compared to
Fig.~\ref{fig-tang}.  Especially the locations of the arms are
unchanged. The inner arms now contain more particles, because of the
initial exponential surface density profile. However, the
line--of--sight velocities have been modified somewhat. For example,
the 3-kpc-arm has now about the right velocity: $\simeq -41\kms$ at
$l=0$. At around $l=20\deg$ the terminal velocities are now larger
than the observed velocities. On the southern side, at $l\sim-20\deg$,
the velocities are also higher, removing part of the previous
discrepancy between model and observation. The larger velocities in
this region are probably due to the mass in the 3-kpc-arm and the
second arm at $\sim2\kpc$ and their symmetric counterarms.

We have also run a model with about twice as much mass in the `arm'
potential and half the smoothing length, $\varepsilon=0.5\kpc$ (right
panels of Fig.~\ref{fig-grav}). Both parameter increase the
gravitational response to the arms and are extreme values for the
Milky Way.  In this case, the distribution of the gas has changed more
dramatically, especially in the region close the the bar. The inner
arms seem to form an ellipse around the bar; however, the number of
arms has still not changed. The 3-kpc-arm in this model is faster than
the observed 3-kpc-arm and expands with $60\kms$ towards the observer
at $l=0$, but it no longer extends out to $3\kpc$. The model terminal
curve now gives only a poor fit to the observations. One interesting
property of this model is that it contains much more gas with
forbidden velocities than all our other models.

From both models it is clear that the gravitational potential of the
spiral arms is important for the comparison with Galactic \lvplot s at
a level of $\sim 15\kms$. Especially the 3-kpc-arm and details in the
terminal velocity curve depend on this parameter. Besides that,
however, the morphology and spiral arm tangents are not effected
much. It is encouraging that the inclusion of spiral arm gravity
appears to improve the fit to some aspects of the data, even though
the models discussed in this section certainly do not contain the
entire story. For example, we have not subtracted a fraction of the
mass in the minor axis NIR clumps which should now be taken care of by
the spiral arms. On a more fundamental level, it is quite possible
that the spiral arms between the bar and the solar circle rotate with
a somewhat different pattern speed from that of the bar (see Sellwood
\& Wilkinson 1993 for a review of this subject, and Amaral \& L\'epine
1997 for a separate model of the Galactic spiral arms with a slow
pattern speed).  If this were the case, we would have to observe the
resulting time--dependent pattern of bar and spiral arms at a moment
when both the apparent NIR distribution of light and the induced
kinematical perturbations resemble those observed in the Milky
Way. This may not be as difficult as it appears because of the Galaxy's
relatively tightly wound four--armed spiral pattern. However,
simulating this would introduce an entirely new degree of freedom and
will not be attempted here.

\section{Conclusions}

We have presented a new set of hydrodynamical models for the gas flow
in the Galactic disk inside the solar circle. These gas flows are
evolved in the gravitational potentials obtained by deprojecting the
NIR luminosity distribution of the Galactic bar and disk from
COBE/DIRBE under the assumption of eight--fold symmetry (as in Binney,
Gerhard \& Spergel 1997), assuming constant mass--to--light ratio
for the NIR luminous material, and adding a nucleus and (in some
cases) a dark halo component. These models allow us to understand
many features of Galactic HI and CO observations.

To follow the gas dynamics we have used the Smooth Particle
Hydrodynamics (SPH) code described in Englmaier \& Gerhard
(1997). With the SPH method we can resolve the spiral arm shocks well,
and use them in some models as tracers for the gravitational potential
of the stellar spiral arms.  The hydrodynamical models quickly settle
to an approximate quasi--equilibrium flow pattern whose overall
morphology is not sensitive to the precise value of the bar's pattern
speed, to the orientation of the bar with respect to the observer, and
to whether or not the spiral arms carry mass.

We have compared our gas models with Galactic HI, CO, HII, and other
data. We find that these models provide a coherent explanation of many
aspects of the data, such as: (i) the four--armed spiral structure of
the Milky Way between corotation and the solar radius, (ii) the nature
of the 3--kpc--arm, (iii) the terminal velocity curve, (iv) the
non--circular velocities near the cusped orbit at the ILR, and (v) the
disk of gas on the inner $x_2$--orbits. Thus NIR photometry and gas
kinematic observations conform to a single picture, and the Galactic
bar is an essential part of this.

In this picture, the bar (bulge) rotates with a pattern speed such
that corotation is at $R_c\simeq3.5\pm 0.5\kpc$. The 3-kpc arm is one
of the arms emanating from the ends of the bar, extending into the
corotation region. Outside corotation, a four--armed spiral arm
pattern gives rise to the molecular ring and the arms extending to the
solar circle and beyond.  In the model, this pattern is generated by
the rotating luminosity/mass concentrations on the bar's minor axis
found by BGS. These can therefore not just be due to light from young
supergiant stars, but must be massive; most likely they are
symmetrized approximations to the stellar spiral arms themselves.
A spiral pattern similar to that found here has been observed by Fux
(1998) in those of his N--body -- SPH barred galaxy models which
compare best to the Galactic \lvplot s. These models start from
a set of specified initial conditions rather than from observations.
The fact that both approaches lead to similar overall results is
encouraging.

We find that the Galactic terminal curve out to longitudes
$l\simeq 45\deg$ is consistent with a maximal, constant mass--to--NIR
light disk and bar model.  The inferred mass in the disk cannot easily
be reduced because (i) this model still underpredicts the microlensing
optical depth towards the bulge (Bissantz \etal 1997) and (ii) it
predicts about the correct surface mass density for the old stellar
disk near the Sun.  Thus the Galactic dark halo will be an important
contribution to the mass of the Milky Way only outside of at least
$R=5\kpc$, depending on the LSR rotation velocity.

The models are similar and in reasonable overall agreement with the
Milky Way observations for a range of values of the bar's orientation
angle. Probably best is $\phibar=20-25\deg$ but the uncertainties are
are such that $15\deg$ or $30\deg$ cannot be excluded.  The detailed
match to various key observational quantities does depend on $\phibar$
and the precise value of $R_c$, but at a level comparable to the
influence of various other parameters including the LSR rotation
velocity, the asymptotic halo circular speed, and the influence of the
stellar spiral arm gravity.

Some features of the HI and molecular emission data can be reproduced
quantitatively, such as the Galactic spiral arm tangent positions,
much of the terminal velocity curve, or the position of the nuclear
$x_2$--disk.  On the other hand, a detailed quantitative fit to all
features in the observed \lvplot s is not yet possible. The 3-kpc arm
in the model has the correct angular extent but somewhat too small
non-circular velocities. Its arm tangent position coincidences with
the place where the model terminal curve differs most from the
observations. In this region the disk mass model may not be very
accurate.  The position and velocity of the molecular parallelogram
are fitted well by closed orbits in the NIR potential, whereas the
hydrodynamic gas flow underestimates the velocity and overestimates
its radial scale. Probably both hydrodynamic resolution and
uncertainties in the potential near $\sim200\pc$ (where the NIR data have
insufficient resolution) are responsible for this. Also, this suggests
that the clouds near the peak of the terminal velocity curve have a
small mean free path.

Much work remains to be done. Because the deprojected disk model
accounts only insufficiently for the Galactic spiral arms, some
aspects of the gravitational potentials used are likely to be
wrong. With a model for the spiral arms in hand, both the correction
of the NIR data for dust and the subsequent deprojection of these data
could be improved. Further observational work on spiral arm tracers such as HII
regions and molecular clouds would be highly valuable for clarifying
the run of the weaker Galactic spiral arms and those on the
other side of the bar, and thus for better constraining the gas-dynamical
models. Distance estimates to these tracers could be improved by
making use of the velocity fields in these models.

On a more fundamental level, several assumptions made in the models
may be or are likely to be invalid at some level, and require further
study: (i) that the gravitational potential and gas flow structures
are quasi-stationary and point--symmetric with respect to the Galactic
Centre; (ii) that the NIR light is a fair tracer of the stellar mass,
i.e., that young supergiant stars do not contribute significantly to
the NIR light (but see Rhoads 1998); (iii) that the gas disk can be
treated as planar; (iv) and that the bar and spiral arms rotate with
the same pattern speed.

\section*{Acknowledgments}

We are grateful to M.~Steinmetz for many discussions on SPH and for
making his original code available. We thank B.~Burton, L.~Bronfman,
T.~Dame and H.~Liszt for sending data, some in advance of publication.
We also thank T.~Dame for producing Figure 1 and M.~Samland for help
with plot software. We acknowledge helpful discussions with J.~Binney,
T.~Dame, R.~Fux and M.~Samland.  This work was supported by the Swiss
NSF grants 21-40'464.94 and 20-43'218.95, and the NASA grants
WKU-522762-98-6 and NAG5-3841.

\section{Multipole expansion}
\label{apx-multipol}

For the SPH simulations, the gravitational forces are calculated
using a multipole expansion of the stellar density model. The stellar
density is given by 
\eq{
	\rho(r,\theta,\varphi) =\rho_0 f(r,\theta,\varphi)
}
where the function $f$ is the interpolated
deprojected light distribution within the bulge region and the
exponential disk elsewhere. Here, the constant $\rho_0$ is the
presently unknown unit of luminosity density multiplied by the also
unknown mass-to-light ratio. A first estimate gives $\rho_0=\xi^2
3\;10^8\;\MSUN/\PC^3$, where $\xi$ is of the order of unity and is
measured by fitting the observed rotation velocity in
\S~\ref{sec-termcurve}.

The density multipoles
\eqn{rholm}{
	\rho_{lm}(r)= \int_0^\pi \!\!{\d}\theta \sin(\theta)
P_{lm}(\cos(\theta))
	       	\int_0^{2\pi} \!\!{\d}\varphi \cos(m\varphi) f({\mathbf r})
}
satisfy the identity equation
\eqn{sumrholm}{
   \rho({\mathbf r})=\sumlm (2-\delta_{m0}){(2l+1)(l-m)!\over4\pi(l+m)!}
	\rho_{lm} P_{lm} \cos(m\varphi).
}
The $P_{lm}$ are the associated Legendre functions. Here we have used
that $f$ is an even function in $\phi$:
$f(r,\theta,\phi)=f(r,\theta,-\phi)$, and have restricted the sum to
positive $m$.

Both integrations were performed with the Romberg method.  The
density multipoles $\rho_{l0}$ were then fitted to a power law $C_{l}
r^{p_{l}}$ in the range of $350\;\PC$ to $500\;\PC$ with the method of
least squares.  The tabulated multipole expansions were then replaced
by the fit inside $350\;\PC$.

From the modified tables $\rho_{lm}(r)$ we calculated the following 
two auxiliary integrals:
\begin{subequations}\label{Iints}
\eqn{Ileft}{
   I_{<}(r) = \int_0^r {\d}a\, \rho_{lm}(a) a^{l+2}
}
and
\eqn{Iright}{
   I_{>}(r) = \int_r^{\infty} {\d}a\, \rho_{lm}(a) a^{1-l}
}
\end{subequations}
For stability reasons we used for these integrations the trapezoidal rule 
on the 3000 logarithmically equidistant $\rho_{lm}$ values
tabulated between $1\;\PC$ and $12\;\KPC$. The region between $12$ and
$16\;\KPC$ in the second integral was calculated again
with the Romberg method; outside $16\;\KPC$ the density was set to zero.

From this, we get the potential multipoles 
\eqn{philm}{
   \Phi_{lm}(r)=(2-\delta_{m0}){(l-m)!\over(l+m)!}
	( r^{-l-1} I_{<}(r) + r^l I_{>}(r) )
}
as well as their first derivatives
\eqn{phidlm}{
   \Phi'_{lm}(r)=(2-\delta_{m0}){(l-m)!\over(l+m)!}
	(-(l+1)r^{-l-2} I_{<}(r)  + l\; r^{l-1} I_{>}(r) ).
}
The potential is then given by
\eqn{sumphilm}{
	\Phi(r)=- G\rho_0 \sumlm \Phi_{lm}(r) P_{lm} \cos(m\varphi).
}

The components of the gravitational acceleration can be
calculated from the partial derivatives of $\Phi$,
\eq{
	{\d\Phi\over\d r}=- G\rho_0 \sumlm \Phi'_{lm} P_{lm}\cos(m\varphi),
}
\eq{
  {\d\Phi\over\d\theta}=G\rho_0 {R\over r} \sumlm \Phi_{lm} 
P'_{lm}\cos(m\varphi),
}
\eq{
	{\d\Phi\over\d\varphi}=G\rho_0 \sumlm m \Phi_{lm} P_{lm} \sin(m\varphi),
}
and the components of the acceleration ${\mathbf a}=-{\mathbf
\nabla}\Phi$ from
\eq{
	a_x = -{\d\Phi\over\d r}x/r + {\d\Phi\over\d\varphi}y/R^2 
		- {\d\Phi\over\d\theta}x z/(r^2 R),
}
\eq{
	a_y = -{\d\Phi\over\d r}y/r - {\d\Phi\over\d\varphi}x/R^2 
		- {\d\Phi\over\d\theta}y z/(r^2 R),
}
\eq{
	a_z = -{\d\Phi\over\d r}z/r + {\d\Phi\over\d\theta} R/r^2.
}

For the gasdynamical model we need only the forces in the galactic plane.
In this special case, it is therefore sufficient to tabulate the
functions
\eq{
	\Phi_m = \suml \Phi_{lm} P_{lm}(0) \Phi_{lm},
}
\eq{
	\Phi'_m = \suml \Phi'_{lm} P_{lm}(0) \Phi_{lm},
}
and to compute the forces using
\eq{
	{\d\Phi\over\d r}=- G\rho_0 \sum_m \Phi'_m \cos(m\varphi),
}
\eq{
	{\d\Phi\over\d\varphi}=G\rho_0 \sum_m m \Phi_m \sin(m\varphi).
}

\end{document}